\documentclass[twocolumn]{autart}    % Enable this line and disable the
\usepackage{tikz}

\usepackage{amsmath}
\usepackage{latexsym, amssymb}
\usepackage{graphicx}
\usepackage{amsmath, amsbsy}
\usepackage{amsopn, amstext}
\usepackage{hyperref}
\usepackage{cancel, color}
\usepackage{epstopdf}

\DeclareMathOperator{\Col}{Col}

\DeclareMathOperator{\lcm}{lcm}
\DeclareMathOperator{\GL}{GL}
\DeclareMathOperator{\id}{id}
\def\cal{\mathcal}

\def\diag{diag}
\def\ra{\rightarrow}

\def\a{\alpha}
\def\b{\beta}
\def\d{\delta}

\def\D{\Delta}

\def\od{\ltimes}

\def\ot{\otimes}

\def\M{\mbox{\boldmath$M$}}

\def\0{{\bf 0}}

\newcommand{\R}{{\mathbb R}}

\def\thefootnote{*}

\newtheorem{dfn}[thm]{Definition}
\newtheorem{prp}[thm]{Proposition}
\newtheorem{exa}[thm]{Example}

\begin{document}

\begin{frontmatter}

\title{Linear Representation of Symmetric Games \thanksref{footnoteinfo}}
% Title, preferably not more

\thanks[footnoteinfo]{This work is supported partly by NNSF
    61333001 and 61273013 of China. Corresponding author: Daizhan Cheng. Tel.: +86 10 6265 1445; fax.: +86 10 6258 7343.}

\author{Daizhan Cheng}\ead{dcheng@iss.ac.cn}, {Ting Liu} %{Yaqi Hao}\dag
%\thanks[CA]{Corresponding author. Tel. +86-10-6265 1445. Fax +86-10-6258 7343.}

\address{Academy of Mathematics and Systems Science,
Chinese Academy of Sciences, Beijing 100190, P.R.China}


\begin{abstract}
Using semi-tensor product of matrices, the structures of several kinds of symmetric games are investigated via the linear representation of symmetric group in the structure vector of games as its representation space. First of all, the symmetry, described as the action of symmetric group on payoff functions, is converted into the product of permutation matrices with structure vectors of payoff functions. Using the linear representation of the symmetric group in structure vectors, the algebraic conditions for the ordinary, weighted, renaming and name-irrelevant symmetries are obtained respectively as the invariance under the corresponding linear representations. Secondly, using the linear representations the relationship between symmetric games and potential games is investigated. This part is mainly focused on Boolean games. An alternative proof is given to show that ordinary, renaming and weighted symmetric Boolean games are also potential ones under our framework. The corresponding potential functions are also obtained. Finally, an example is given to show that some other Boolean games could also be potential games.
\end{abstract}

\end{frontmatter}

\section{Introduction}%s-1

Symmetric game is an important class of games. It has drawn a considerable attention from both game theoretical community and system scientists. There are several reasons for this: First of all, symmetry represents ``fair". In real world, a fair game is more realistic and acceptable. That makes many commonly played games symmetric. For instance, Rock-Paper-Scissors, Prisoner's Dilemma, Hawk and Dove, etc. are all symmetric. If more general types of symmetric games are considered, many other games such as Battle of the sexes, Matching  Pennies, etc. are also symmetric.  Secondly, symmetry can greatly simplify the representation and computation of game related issues \cite{pap08}. Finally, symmetry may be related with certain other properties. For instance, it is well known that a finite symmetric Boolean game, i.e., a game with a common action set of size $2$,  is an exact potential game \cite{hof02}.

The concept of symmetric games was firstly proposed by J. Nash in his famous paper \cite{nas51}. Lately, it has been developed into several classes of interesting games  \cite{alo13}, \cite{kub13}, \cite{cao16}. According to \cite{cao16}, there are two commonly recognized classes of symmetric games: (1) ordinary symmetric game; (2) name-irrelevant symmetric game. The first one is very natural, which means all the payoffs are essentially the same. The second one is firstly proposed by Peleg et al \cite{pel99}, based on a simple type of strategy permutations of Nash  \cite{nas51}. A new class of symmetric games was proposed in \cite{cao16} as the renaming symmetric game. (See also \cite{caopr2}).

From mathematical point of view, a symmetry means invariance under the action of certain group. Roughly speaking, a game is symmetric means its payoff functions are invariant under a permutation group, which is a subgroup of a symmetric group \cite{dix96}.

Linear representation of a group $A$ in a vector space $V$ is an isomorphism $A\ra \GL(V)$ \cite{ser77}. This representation not only makes the group action easily computable but also provides a framework to investigate certain properties of the action. In this paper we reveal for each symmetry of games its linear representation, which is the action of symmetric group or its subgroups on the structure vector of games. The structure vector of a game consists of the structure vectors of its payoff functions.

To construct and to investigate the linear representations of symmetric games the semi-tensor product (STP) of matrices becomes a main tool. STP of matrices is a newly developed matrix product \cite{che12}. It is a generalization of conventional matrix product and keeps most properties of conventional matrix product remaining available. It has been successfully used for studying logical (control) systems. We refer to \cite{che11}, \cite{las13}, \cite{for13}, \cite{li13}, \cite{wan12}, \cite{zou15}, \cite{zho16}, just to name a few. It has also been used to game theoretic control problems  \cite{guo13}, \cite{che14}, \cite{che15}, \cite{zou16}.

Vector space structure of finite games is a key issue in this linear representation approach to symmetric games. A clear picture of vector space structure of finite games  was firstly presented in \cite{can11}. This vector space structure was then modified and merged into an Euclidean space \cite{che16}. In this paper a finite game $G$ with player set $|N|=n$ and strategy set $|S_i|=k_i$, $i=1,\cdots,n$, is simply considered as a point in $\R^s$, where $s=n\prod_{i=1}^nk_i$, denoted by $V_G\in \R^s$. Such a set is denoted by ${\cal G}_{[n;k_1,\cdots,k_n]}$, which has the same topology and vector space structure as $\R^s$, as demonstrated in \cite{che16}.

The first purpose of this paper is to deduce the linear representations for each of the aforementioned symmetries of finite games, considering as the action of symmetric group or its proper subgroups in the vector space ${\cal G}_{[n;k_1,\cdots,k_n]}\sim \R^s$, which consists of the structure vectors of all payoff functions.
Then it is proved that a finite game $G$ is certain kind of symmetry, if and only if, its structure vector is invariant under the corresponding linear representation. This presentation gives an easy way to verify whether a finite game $G$ is of this kind of symmetry. Moreover, the matrix form of linear representations provides a framework for investigating the properties of symmetric games and manipulating them.

Secondly, the relationship between certain symmetric games and potential games is investigated. We mainly focus on Boolean games. The ordinary and renaming symmetric Boolean games are proved to be potential games and the weighted symmetric Boolean games are proved to be weighted potential games.

Thirdly, a special kind of Boolean games, called the negation-symmetric Boolean game, is defined and investigated. We prove that this kind of Boolean games is also potential. It shows that (generalized) ordinary symmetry is sufficient for a Boolean game to be potential, but it is not necessary.

Finally, for statement ease, we give some notations:

\begin{enumerate}

\item  ${\cal M}_{m\times n}$: the set of $m\times n$ real matrices.

\item ${\cal B}_{m\times n}$: the set of $m\times n$ Boolean matrices, (${\cal B}_{n}$: the set of $n$ dimensional Boolean vectors.)

%\item $\Col(M)$ ($\Row(M)$): the set of columns (rows) of $M$. $\Col_i(M)$ ($\Row_i(M)$): the $i$-th column (row) of $M$.

\item ${\cal D}:=\left\{0,1\right\}$.

\item $\d_n^i$: the $i$-th column of the identity matrix $I_n$.

\item $\D_n:=\left\{\d_n^i\vert i=1,\cdots,n\right\}$.

\item ${\bf 1}_{\ell}=(\underbrace{1,1,\cdots,1}_{\ell})^T$.

\item ${\bf 0}_{p\times q}$: a $p\times q$ matrix with zero entries.

\item A matrix $L\in {\cal M}_{m\times n}$ is called a logical matrix
if the columns of $L$ are of the form of
$\d_m^k$. That is, $\Col(L)\subset \D_m$.
Denote by ${\cal L}_{m\times n}$ the set of $m\times n$ logical
matrices.

\item If $L\in {\cal L}_{n\times r}$, by definition it can be expressed as
$L=[\d_n^{i_1},\d_n^{i_2},\cdots,\d_n^{i_r}]$. For the sake of
compactness, it is briefly denoted as $
L=\d_n[i_1,i_2,\cdots,i_r]$.

\item $M_n$, the structure matrix of ``negation", that is, $M_n=\d_2[2,1]$.
\item ${\bf S}_n$: $n$-th order symmetric group.
\item ${\bf P}_n$: $n$-th order Boolean orthogonal group.
\item Let $H,~G$ be two groups. $H<G$ means $H$ is a subgroup of $G$.
\item ${\bf \Theta}_{[n;\kappa]}$: block permutation group (${\bf \Theta}_{[n;\kappa]}< {\bf S}_{n\kappa}$).
\item $\id$: identity mapping.
\item $\GL(n,\R)$ (or~$\GL(V)$): general linear group.
\item ${\cal G}_{[n;k_1,\cdots,k_n]}$: Set of finite games with $|N|=n$, $|S_i|=k_i$.
\item ${\cal G}_{[n;\kappa]}$: $|S_i|=\kappa$, $i=1,\cdots,n$.
\item ${\cal S}^o_{[n;\kappa]}$: the set of ordinary symmetric games.
\item ${\cal S}^w_{[n;\kappa]}$: the set of weighted symmetric games.
\item ${\cal S}^r_{[n;\kappa]}$: the set of renaming symmetric games.
\end{enumerate}

The rest of this paper is organized as follows: Section 2 considers the action of symmetric group on games. Subsection 2.1 introduces symmetric group and its matrix expression. Subsection 2.2 reviews the semi-tensor product (STP) of matrices, which is a fundamental tool in this paper. Using STP, Subsection 2.3 expresses the group action on finite game into matrix form. Section 3 provides the linear representation for ordinary symmetry of games. Continued from Section 3, Section 4 introduces the weighted symmetry and renaming symmetry, which are two simple extensions of ordinary symmetry. The name-irrelevant symmetry and its linear representation are discussed in Section 5. Some examples are presented to depict them. The potential game is introduced in Section 6. The verification of potential games and the formula for calculating potential function are also presented. Section 7 shows the relationship between potential games and symmetric games for Boolean games only. It was shown that an ordinary symmetric Boolean game and a renaming symmetric Boolean game are potential game. (These conclusions are well known, but we give an alternative proof.) And a weighted symmetric Boolean game is a weighted potential game. The corresponding potential functions are also obtained respectively. Section 8 introduces the negation-symmetric Boolean game and proved that it is also a potential game. Section 9 is a concluding remark, which shows that all the symmetries discussed in this paper can be characterized by a unified form of linear representations. This characteristic property reveals the relationship among them. The relationship between symmetric games and potential games is also summarized. To streamline the presentation, some proofs are provided in the Appendix.

\section{Action of Symmetric Group on Games}%s-2

\subsection{Symmetric Group and Boolean Orthogonal Group}%s-2.1

Let $N=\{1,2,\cdots,n\}$. ${\bf S}_n$ is the set of permutations of elements of $N$. With the composed permutation as the product of two permutations, ${\bf S}_n$ becomes a group. The elements in ${\bf S}_n$ can be expressed either by a one-one correspondence or a product of cycles. We use a simple example to depict them.

\begin{exa}\label{e2.1.1}
Consider ${\bf S}_5$, $\sigma,~\mu \in {\bf S}_5$ can be expressed as
$$
\sigma:~\begin{array}{ccccc}
1&2&3&4&5\\
\downarrow&\downarrow&\downarrow&\downarrow&\downarrow\\
3&5&1&4&2
\end{array};\quad \mu:~\begin{array}{ccccc}
1&2&3&4&5\\
\downarrow&\downarrow&\downarrow&\downarrow&\downarrow\\
2&4&5&3&1
\end{array}.
$$
Equivalently, they can be expressed in cycle form as
$$
\sigma=(1,3)(2,5);\quad \mu=(1,2,4,3,5).
$$
The product of $\sigma$ and $\mu$ is defined as
$$
\mu\circ \sigma=(1,5,4,3,2).
$$
Note that in the above product, the permutation $\sigma$ is performed first, and then $\mu$ is performed. Say,
$1\xrightarrow{\sigma}3\xrightarrow{\mu} 5$, which leads to $1\xrightarrow{\mu\circ\sigma} 5$,
etc.
\end{exa}

Let $G$ be a group and $X\neq \emptyset$, $\phi:G\times X\ra X$ is called a group action of $G$ on $X$, if
$$
\begin{array}{l}
\phi(e,x)=x,\quad e\in G~\mbox{is the identity},~x\in X\\
\phi(g_2,\phi(g_1,x))=\phi(g_2g_1,x),\quad g_1,~g_2\in G,~x\in X.
\end{array}
$$

The symmetric group has a natural action on $N$, that is, $\sigma\times i\mapsto \sigma(i)$.

For each $\sigma\in {\bf S}_n$, we can construct a Boolean orthogonal matrix, called the structure matrix of $\sigma$, as
\begin{align}\label{2.1.1}
P_{\sigma}=\left[\d_n^{\sigma(1)},\d_n^{\sigma(2)},\cdots,\d_n^{\sigma(n)}\right].
\end{align}

Define the set of structure matrices of $n$-th order permutations as
\begin{align}\label{2.1.2}
{\bf P}_n:=\left\{P_{\sigma}\;|\;\sigma\in {\bf S}_n\right\}.
\end{align}
Note that the set of $n$-th order Boolean orthogonal matrices is a subgroup of the general linear group $\GL(n,\R)$. Moreover, it is obvious that there is a one-one correspondence between ${\bf P}_n$ and the set of $n$-th order Boolean orthogonal matrices. We, therefore, have ${\bf P}_n< \GL(n,\R)$. Next, define $\psi:{\bf S}_n\ra {\bf P}_n$ as
\begin{align}\label{2.1.3}
\psi(\sigma):=P_{\sigma},
\end{align}
then $\psi$ is a group isomorphism. That is, $\psi$ is bijective and
\begin{align}\label{2.1.4}
\psi(\mu\circ \sigma)=P_{\mu\circ \sigma}=P_{\mu}P_{\sigma}=\psi(\mu)\psi(\sigma).
\end{align}

It is easy to see that
\begin{enumerate}
 \item ${\bf P}_n$ has a natural group action on $\D_n$ as
\begin{align}\label{2.1.5}
P_{\sigma}\d_n^i=\d_n^{\sigma(i)}\in \D_n,\quad i=1,\cdots,n;
\end{align}
\item if we identify $i\sim \d_n^i$, then the group action of ${\bf S}_n$ on $N$ is exactly the same as the group action of ${\bf P}_n$ on $\D_n$. Precisely,
\begin{align}\label{2.1.6}
\sigma(i)\sim P_{\sigma}\d_n^i.
\end{align}
\end{enumerate}

\subsection{Semi-tensor Product of Matrices}%s-2.2

This subsection is a brief introduction to semi-tensor product (STP) of matrices. We refer to \cite{che11}, \cite{che12} for details. The STP of matrices is defined as follows:

\begin{dfn}\label{d2.2.1}  Let $M\in {\cal M}_{m\times n}$, $N\in {\cal M}_{p\times q}$, and $t=\lcm\{n,p\}$ be the least common multiple of $n$ and $p$.
The STP of $M$ and $N$ is defined as
\begin{align}\label{2.2.1}
M\ltimes N:= \left(M\otimes I_{t/n}\right)\left(N\otimes I_{t/p}\right)\in {\cal M}_{mt/n\times qt/p},
\end{align}
where $\otimes$ is the Kronecker product.
\end{dfn}

 The STP of matrices is a generalization of conventional matrix product, and all the computational properties of the  conventional matrix product remain available. Throughout this paper, the default matrix product is STP, so the product of two arbitrary matrices is well defined, and the symbol $\ltimes$ is mostly omitted.

First, we give some basic properties of STP, which will be used in the sequel.

 \begin{prp} \label{p2.2.2}
 \begin{enumerate}
 \item (Associative Law:)
 \begin{align}\label{2.2.2}
 A\ltimes (B\ltimes C)=(A\ltimes B)\ltimes C.
 \end{align}
 \item (Distributive Law:)
 \begin{align}\label{2.2.3}
 (A+B)\ltimes C=A\ltimes C+B\ltimes C;
 \end{align}
 \begin{align}\label{2.2.4}
 C\ltimes (A+B)=C\ltimes A+C\ltimes B.
 \end{align}
 \end{enumerate}
\end{prp}

 \begin{prp} \label{p2.2.3} Let $X\in \R^t$ be a $t$ dimensional column vector, and $M$ a matrix. Then
\begin{align}\label{2.2.5}
X\ltimes M=\left(I_t\otimes M\right)\ltimes X.
\end{align}
\end{prp}

Next, we consider the matrix expression of logical relations. Identifying
$$
1\sim \d_2^1,\quad 0\sim \d_2^2,
$$
then a logical variable $x\in {\cal D}$ can be expressed in vector form as
$$
x\sim\begin{pmatrix}x\\1-x\end{pmatrix},
$$
which is called the vector form expression of $x$

A mapping $f:{\cal D}^n\ra \R$ is called a pseudo-Boolean function.

\begin{prp} \label{p2.2.4} Given a pseudo-Boolean function $f:{\cal D}^n\ra \R$,  there exists a unique row vector $V_f\in \R^{2^n}$, called the structure vector of $f$,  such that (in vector form)
\begin{align}\label{2.2.6}
f(x_1,\cdots,x_n)=V_f\ltimes_{i=1}^nx_i.
\end{align}
\end{prp}

\begin{rem}\label{r2.2.401} In previous proposition, if ${\cal D}$ is replaced by ${\cal D}_k$, $k>2$, then the function $f$ is called a pseudo-logical function and the result remains available with an obvious modification that $x_i\in \D_k$ and $V_f\in \R^{k^n}$.
\end{rem}

\begin{dfn}\label{d2.2.5} A swap matrix $W_{[m,n]}\in {\cal M}_{mn\times mn}$ is defined as
\begin{align}\label{2.2.7}
\begin{array}{ccl}
W_{[m,n]}&:=&[\d_n^1\d_m^1,\d_n^2\d_m^1,\cdots,\d_n^n\d_m^1,\d_n^1\d_m^2,\\
~&~&~~\cdots,\d_n^n\d_m^2,\cdots,
\d_n^1\d_m^m,\cdots,\d_n^n\d_m^m].
\end{array}
\end{align}
\end{dfn}

The basic function of a swap matrix is to swap two vectors.

\begin{prp}\label{p2.2.6} Let $X\in \R^m$ and $Y\in \R^n$ be two column vectors. Then
\begin{align}\label{2.2.8}
W_{[m,n]}XY=YX.
\end{align}
\end{prp}

The swap matrix has following decomposition property.

\begin{prp}\label{p2.2.7}
\begin{align}\label{2.2.9}
W_{[p,qr]}=\left(I_q\otimes W_{[p,r]}\right)\left(W_{[p,q]}\otimes I_r\right).
\end{align}
\end{prp}
%
%Define a power reducing matrix as
%\begin{align}\label{2.2.10}
%R^P_k:=\diag\left(\d_k^1,\d_k^2,\cdots,\d_k^k\right)\in {\cal B}_{k^2\times k}.
%\end{align}
%It can be used to reduce the power of logical variables.
%
%\begin{prp}\label{p2.2.8} Assume $x\in \D_k$, then
%\begin{align}\label{2.2.11}
%x\ltimes x=R^P_kx.
%\end{align}
%\end{prp}

\begin{dfn}\label{d2.2.9} \cite{kha68,che12} Assume $A\in {\cal M}_{p\times s}$, $B\in {\cal M}_{q\otimes s}$, the Khatri-Rao product of $A,~B$, denoted by $A*B$, is defined as
\begin{align}\label{2.2.12}
\begin{array}{ccl}
A*B&=&\left[\Col_1(A)\ltimes \Col_1(B), \Col_2(A)\ltimes \Col_2(B),\right.\\
~&~&\left.\cdots,\Col_s(A)\ltimes \Col_s(B)\right]\in {\cal M}_{pq\times s}.
\end{array}
\end{align}
\end{dfn}

\begin{prp}\label{p2.2.10} Assume $x\in \D_n$, $y\in \D_p$, $z\in \D_q$, where $y,~z$ are logical functions of $x$ and are expressed in algebraic form as
\begin{align}\label{2.2.13}
\begin{array}{ccl}
y=Mx,\quad M\in {\cal L}_{p\times n}\\
z=Nx,\quad N\in {\cal L}_{q\times n}.
\end{array}
\end{align}
Then we have
\begin{align}\label{2.2.14}
yz=(M*N)x.
\end{align}
\end{prp}

\subsection{Matrix Expression of Finite Games under Action of ${\bf S}_n$}%s-2.3

A finite game, denoted by a triple $G=(N,S,C)$, consists of three components: (i)  the set of players $N=\{1,2,\cdots,n\}$; (ii) the profile set $S=\prod_{i=1}^nS_i$, where $S_i=\{1,2,\cdots,k_i\}$, is the set of strategies of player $i$, $i=1,\cdots,n$; the set of payoffs $C=\{c_1,\cdots,c_n\}$, where $c_i:S\ra \R$ is the payoff function of  player $i$, $i=1,\cdots,n$. Denote the set of $G$ with $|N|=n$ and $|S_i|=k_i$ by ${\cal G}_{[n;k_1,\cdots,k_n]}$. If $|S_i|=\kappa$, $i=1,\cdots,n$, the set of such games is briefly denoted by ${\cal G}_{[n;\kappa]}$. This paper concerns only ${\cal G}_{[n;\kappa]}$.

Assume a game $G\in {\cal G}_{[n;\kappa]}$ is given. It is obvious that the payoff function $c_i$ is a pseudo-logical function. When the strategies $x_i$, $i=1,\cdots,n$, are expressed into their vector form, the payoff functions can be expressed as
\begin{align}\label{2.3.1}
c_i(x_1,\cdots,x_n)=V^c_i\ltimes_{j=1}^nx_j:=V^c_ix,\quad i=1,\cdots,n,
\end{align}
where
\begin{align}\label{2.3.2}
x=\ltimes_{i=1}^nx_i\in \D_{\kappa^n}
\end{align}
is called the STP form of the strategy profile. We can also express it in a (column) vector form as
\begin{align}\label{2.3.3}
\vec{x}:=[x_1^T,x_2^T,\cdots,x_n^T]^T\in {\cal B}_{n\kappa}.
\end{align}
Denoting
\begin{align}\label{2.3.4}
\Phi_i:={\bf 1}^T_{\kappa^{i-1}}\otimes I_{\kappa}\otimes {\bf 1}^T_{\kappa^{n-i}},\quad i=1,\cdots,n,
\end{align}
then a matrix $\Phi$ can be constructed as
\begin{align}\label{2.3.5}
\Phi:=\begin{bmatrix}
\Phi_1\\\Phi_2\\\vdots\\\Phi_n
\end{bmatrix}\in {\cal B}_{n\kappa\times \kappa^n}.
\end{align}
This $\Phi$ can convert a profile from its STP form into its vector form.

\begin{prp}\label{p2.3.1} Let $G\in {\cal G}_{[n;\kappa]}$, $x$ and $\vec{x}$ are STP form and vector form of its strategy profile respectively. Then
\begin{align}\label{2.3.6}
\vec{x}=\Phi x.
\end{align}
\end{prp}

Next, assume $\sigma\in {\bf S}_n$. A straightforward computation shows that
\begin{align}\label{2.3.7}
P_{\sigma}\vec{x}=\begin{bmatrix}
x_{\sigma^{-1}(1)}\\
x_{\sigma^{-1}(2)}\\
\vdots\\
x_{\sigma^{-1}(n)}\\
\end{bmatrix}=\begin{bmatrix}
\Phi_{\sigma^{-1}(1)}\\
\Phi_{\sigma^{-1}(2)}\\
\vdots\\
\Phi_{\sigma^{-1}(n)}\\
\end{bmatrix}x.
\end{align}

The following expression can be proved by using formula (\ref{2.3.7}) and Proposition \ref{p2.2.10}.

\begin{prp}\label{p2.3.2} Let $\sigma\in {\bf S}_n$.
Define
\begin{align}\label{2.3.8}
\sigma(x):=\ltimes_{i=1}^nx_{\sigma^{-1}(i)}.
\end{align}
Then
\begin{align}\label{2.3.9}
\sigma(x)=T_{\sigma}x,
\end{align}
where
\begin{align}\label{2.3.10}
T_{\sigma}=\Phi_{\sigma^{-1}(1)}*\Phi_{\sigma^{-1}(2)}*\cdots* \Phi_{\sigma^{-1}(n)}.
\end{align}
\end{prp}

Finally, we give an expression for calculation $\Phi_{\sigma^{-1}(i)}$. In fact, substituting (\ref{2.3.6}) into (\ref{2.3.7}) and we have
\begin{align}\label{2.3.11}
\begin{bmatrix}
\Phi_{\sigma^{-1}(1)}\\
\Phi_{\sigma^{-1}(2)}\\
\vdots\\
\Phi_{\sigma^{-1}(n)}\\
\end{bmatrix}=P_{\sigma}
\begin{bmatrix}
\Phi_{1}\\
\Phi_{2}\\
\vdots\\
\Phi_{n}\\
\end{bmatrix}.
\end{align}

\section{Ordinary Symmetry and its Linear Representation}%s_3

Define the structure vector of $G\in {\cal G}_{[n;\kappa]}$ as
\begin{align}\label{2.4.1}
V_G:=[V^c_1,V^c_2,\cdots,V^c_n]\in \R^{n\kappa^n},
\end{align}
then the game $G$ is completely determined by $V_G$. In this way, we have a vector space structure (precisely, $\R^{n\kappa^n}$) for ${\cal G}_{[n;\kappa]}$. We refer to \cite{can11}, \cite{che16} for the vector structure of finite games.

Hereafter, we adopt the terminologies used in \cite{cao16} to describe the symmetry of games.

\begin{dfn}\label{d2.4.1} A game $G\in {\cal G}_{[n;\kappa]}$ is called ordinary symmetric if for any $\sigma\in {\bf S}_n$
\begin{align}\label{2.4.2}
\begin{array}{ccl}
c_i(x_1,\cdots,x_n)&=&c_{\sigma(i)}(x_{\sigma^{-1}(1)}, x_{\sigma^{-1}(2)},\cdots,x_{\sigma^{-1}(n)}),\\
~&~& i=1,\cdots,n.
\end{array}
\end{align}
\end{dfn}

Using (\ref{2.3.7})-(\ref{2.3.9}), we have the following proposition.

\begin{prp}\label{p2.4.2}  A game $G\in {\cal G}_{[n;\kappa]}$ is ordinary symmetric, if and only if, for any $\sigma\in {\bf S}_n$
\begin{align}\label{2.4.3}
V^c_i=V^c_{\sigma(i)}T_{\sigma},\quad i=1,\cdots,n.
\end{align}
\end{prp}

Denote
$$
V_G^{\sigma}:=[V^c_{\sigma(1)},\cdots,V^c_{\sigma(n)}].
$$
Then we have the following result:

\begin{thm}\label{t2.4.3} $G\in {\cal G}_{[n;\kappa]}$ is ordinary symmetric, if and only if,
\begin{align}\label{2.4.4}
V_G(P_{\sigma}\otimes T_{\sigma})=V_G,\quad \forall \sigma\in {\bf S}_n.
\end{align}
\end{thm}

\noindent{\it Proof}. According to Proposition \ref{p2.4.2}, it is enough to show that (\ref{2.4.4}) and (\ref{2.4.3}) are equivalent.
From (\ref{2.4.3}), we have the following identity
\begin{align}\label{2.4.5}
V_G=[V_{\sigma(1)}^c T_{\sigma},V_{\sigma(2)}^c T_{\sigma},\cdots, V_{\sigma(n)}^c T_{\sigma}]
= V_G^{\sigma} (I_n\ot T_{\sigma})
\end{align}
On the other hand, a straightforward computation verifies
\begin{align}\label{2.4.6}
V_G^{\sigma}=V_GP_{\sigma}.
\end{align}

Substituting it into (\ref{2.4.5}) gives rise to (\ref{2.4.4}). On the contrary, splitting (\ref{2.4.4}) to each block component $V^c_i$, it is easy to obtain (\ref{2.4.5}) which yields (\ref{2.4.3}). This completes the proof of Theorem \ref{t2.4.3}.
\hfill $\Box$

\begin{dfn}\label{d2.4.4} \cite{ser77}
A linear representation of a finite group $A$ in a finite vector space $V$ is a homomorphism $\psi$ from the group $A$ into the group of $\GL(V)$.
\end{dfn}

Now consider the vector space $ {\cal G}_{[n;\kappa]}\sim \R^{n\kappa^n}$. That is, each $G\in {\cal G}_{[n;\kappa]}$ is represented by its structure vector $V_G$. Then we have a linear representation of $\sigma\in {\bf S}_{n}$ in  ${\cal G}_{[n;\kappa]}$ as follows:

\begin{prp}\label{p2.4.5} Define a mapping of $\sigma\in {\bf S}_{n}$ in  vector space ${\cal G}_{[n;\kappa]}$ as
\begin{align}\label{2.4.7}
\varphi(\sigma):=P_{\sigma}\otimes T_{\sigma}\in \GL( {\cal G}_{[n;\kappa]}).
\end{align}
Then $\varphi$ is a linear representation of ${\bf S}_n$ in $ {\cal G}_{[n;\kappa]}$.
\end{prp}

\noindent{\it Proof}.
We already have (\ref{2.1.4}), that is,
$$
P_{\mu\circ \sigma}=P_{\mu}P_{\sigma}.
$$
What do we need to prove is
\begin{align}\label{2.4.8}
T_{\mu\circ \sigma}=T_{\mu}T_{\sigma}.
\end{align}
This equation comes from (\ref{2.3.10}) and (\ref{2.3.11}).

\hfill $\Box$

\begin{cor}\label{c2.4.6} $G\in {\cal G}_{[n;\kappa]}$ is ordinary symmetric, if and only if, it is invariant with respect to the linear representation $\varphi$, defined by (\ref{2.4.7}).
\end{cor}

We give an example to depict this:

\begin{exa}\label{e2.4.7} Consider $G\in {\cal G}_{[3;3]}$. Denote
$$
{\bf S}_3=\{\sigma_i\;|\;i=1,2,3,4,5,6\},
$$
where
$$
\begin{array}{lll}
\sigma_1=\id&\sigma_2=(2,3)&\sigma_3=(1,2)\\
\sigma_4=(1,2,3)&\sigma_5=(1,3,2)&\sigma_6=(1,3).\\
\end{array}
$$
Then we have
$$
\begin{array}{ll}
P_{\sigma_1}=I_3&
P_{\sigma_2}=\begin{bmatrix}
     1&     0&     0\\
     0&     0&     1\\
     0&     1&     0
\end{bmatrix}\\
P_{\sigma_3} =\begin{bmatrix}
     0&     1&     0\\
     1&     0&     0\\
     0&     0&     1
\end{bmatrix}&
P_{\sigma_4} =\begin{bmatrix}
     0&     0&     1\\
     1&     0&     0\\
     0&     1&     0
\end{bmatrix}\\
P_{\sigma_5} =\begin{bmatrix}
     0&     1&     0\\
     0&     0&     1\\
     1&     0&     0
\end{bmatrix}&
P_{\sigma_6} =\begin{bmatrix}
     0&     0&     1\\
     0&     1&     0\\
     1&     0&     0
\end{bmatrix};
\end{array}
$$
and
$$
\begin{array}{ccl}
T_{\sigma_1}&=&\id\\
T_{\sigma_2}&=&\d_{27}[ 1,  4,  7,  2,  5,  8,  3,  6,  9,\\
  ~&~&~~~10,  13,  16,  11,  14,  17,  12,  15,  18,\\
  ~&~&~~~19,  22,  25,  20,  23,  26,  21,  24,  27]\\
T_{\sigma_3}&=&\d_{27}[ 1,  2,  3,  10,  11,  12,  19,  20,  21,\\
~&~&~~~4,  5,  6,  13,  14,  15,  22,  23,  24,\\
~&~&~~~  7,  8,  9,  16,  17,  18,  25,  26, 27 ]\\
T_{\sigma_4}&=&\d_{27}[ 1,  10,  19,  2,  11,  20,  3,  12,  21,\\
~&~&~~~4,  13,  22,  5,  14,  23,  6,  15,  24,\\
~&~&~~~  7,  16,  25,  8,  17,  26,  9,  18,  27 ]\\
T_{\sigma_5}&=&\d_{27}[ 1,  4,  7,  10,  13,  16,  19,  22,  25,\\
  ~&~&~~~2,  5,  8,  11,  14,  17,  20,  23,  26,\\
  ~&~&~~~  3,  6,  9,  12,  15,  18,  21,  24,  27 ]\\
T_{\sigma_6}&=&\d_{27}[1,  10,  19,  4,  13,  22,  7,  16,  25,\\
  ~&~&~~~2,  11,  20,  5, 14,  23,  8,  17,  26,\\
  ~&~&~~~  3,  12,  21,  6,  15,  24,  9,  18,  27].
  \end{array}
$$
According to Theorem \ref{t2.4.3}, $G$ is ordinary symmetric, if and only if,
\begin{align}\label{2.4.9}
V_G^T\in \bigcap_{i=1}^6\left(P^T_{\sigma_i}\otimes T^T_{\sigma_i}-I_{81}\right)^{\perp}.
\end{align}
A simple computation shows that (\ref{2.4.9}) implies that
\begin{align}\label{2.4.10}
\begin{array}{ccl}
V_G&=&[A,D,E,D,F,G,E,G,H\\
~&~&I,J,K,J,B,L,K,L,M\\
~&~&N,O,P,O,Q,R,P,R,C\\
~&~&A,D,E,I,J,K,N,O,P\\
~&~&D,F,G,J,B,L,O,Q,R\\
~&~&E,G,H,K,L,M,P,R,C\\
~&~&A,I,N,D,J,O,E,K,P\\
~&~&D,J,O,F,B,Q,G,L,R\\
~&~&E,K,P,G,L,R,H,M,C],
\end{array}
\end{align}
where the parameters $A,B,\cdots,R$ are arbitrary real numbers. Denote by ${\cal S}^o_{[n;\kappa]}$ the ordinary symmetric subspace of ${\cal G}_{[n;\kappa]}$. Then we know that
$$
\dim\left({\cal S}^o_{[3;3]}\right)=18,
$$
and a basis of ${\cal S}^o_{[3;3]}$ can easily be figured out from (\ref{2.4.10}).
\end{exa}

\section{Generalized Ordinary Symmetric Game}%s-4

\subsection{Weighted Symmetric Games}%s-4.1

\begin{dfn}\label{d3.1.1} Let $G\in {\cal G}_{[n;\kappa]}$, $\mu_i\in \R_+$, $i=1,\cdots,n$, be a set of positive real numbers, called the weights.
\begin{enumerate}
\item $G$ is said to be weighted symmetric with respect to $\sigma\in {\bf S}_n$ and $\{\mu_i, i=1,\cdots,n\}$, if
\begin{align}\label{3.1.1}
\begin{array}{l}
\mu_ic_i(x_1,\cdots,x_n)=\mu_{\sigma(i)}c_{\sigma(i)}\left(x_{\sigma^{-1}(1)}, \cdots,\right.\\
~~~\left.x_{\sigma^{-1}(n)}\right),\quad i=1,\cdots,n.
\end{array}
\end{align}
\item $G$ is said to be weighted symmetric with respect to $\{\mu_i,i=1,\cdots,n\}$ if it is weighted symmetric with respect to each $\sigma\in {\bf S}_n$ and $\mu_i$, $i=1,\cdots,n$.
\end{enumerate}
\end{dfn}

\begin{rem}\label{r3.1.2} When $\mu_i=1$, $i=1,\cdots,n$, the weighted symmetry becomes the classical ordinary symmetry.
\end{rem}

\begin{exa}\label{e3.1.3} Consider $G\in {\cal G}_{[2;2]}$. The payoff bi-matrix is in Table \ref{Tab2.1}.
\begin{table}[!htbp] %Default value is [tbp]
\centering \caption{Payoff Bi-matrix \label{Tab2.1}}
%\vskip 2mm
\doublerulesep 0.5pt
\begin{tabular}{|c||c|c|}
\hline
\hline $P_1 \backslash P_2$&$1$&$2$\\
\hline $1$& $2,~3$&$4,~9$\\
\hline $2$&$6,~6$&$4,~6$ \\
\hline
\end{tabular}
\end{table}

Then
$$
\begin{array}{ll}
3c_1(1,1)=2c_2(1,1)=6&3c_1(2,2)=2c_2(2,2)=12\\
3c_1(1,2)=2c_2(2,1)=12&3c_1(2,1)=2c_2(1,2)=18.\\
\end{array}
$$
Hence $G$ is a weighted symmetric game with weights: $\mu_1=3$ and $\mu_2=2$.
\end{exa}

Similar to Proposition \ref{p2.4.2}, we have the following result.

\begin{prp}\label{p3.1.4} $G\in {\cal G}_{[n;\kappa]}$ is weighted symmetric with respect to $\{\mu_i\;|\;i=1,\cdots,n\}$, if and only if, for each $\sigma\in {\bf S}_n$
\begin{align}\label{3.1.2}
\mu_iV^c_i=\mu_{\sigma(i)}V^c_{\sigma(i)}T_{\sigma},
\end{align}
where $T_{\sigma}$ is defined in (\ref{2.3.10}).
\end{prp}

Define a weight matrix $\Gamma$ as
$$
\Gamma:=\diag(\mu_1,\mu_2,\cdots,\mu_n),
$$
where $\mu_i>0$, $i=1,\cdots,n$. Using $\Gamma$, for each $\sigma\in {\bf S}_n$ we define
\begin{align}\label{3.1.3}
\varphi^{\mu}(\sigma):=\Gamma\left(P_{\sigma}\otimes T_{\sigma}\right)\Gamma^{-1}\in \GL({\cal G}_{[n;\kappa]}).
\end{align}
Corresponding to Theorem \ref{t2.4.3}, Proposition \ref{p2.4.5}, and Corollary \ref{c2.4.6}, we have
\begin{thm}\label{t3.1.5}
\begin{enumerate}
\item The mapping $\varphi^{\mu}$ defined in (\ref{3.1.3}) is a linear representation of ${\bf S}_n$ in ${\cal G}_{[n;\kappa]}$.
\item A game $G\in {\cal G}_{[n;\kappa]}$ is weighted ordinary symmetric with respect to $\{\mu_i,\: i=1,\cdots,n\}$, if and only if, $V_G$ is invariant with respect to this linear representation.
\end{enumerate}
\end{thm}

\subsection{Renaming Symmetric Game}%s-4.2

\begin{dfn}\label{d3.2.1} \cite{cao16}
Let $G\in {\cal G}_{[n;\kappa]}$, then
\begin{enumerate}
\item A renaming of $G$ is a set of permutations $r_i\in {\bf S}_{\kappa}$ on $S_i$, $i=1,\cdots,n$.
\item A renaming game $G^r:= \left(N, (S^r_i)_{i\in N}, (c^r_i)_{i\in N}\right)$, where
\begin{align}\label{3.2.1}
\begin{array}{l}
S^r_i=\left\{r_i(1),r_i(2),\cdots,r_i(\kappa)\right\},\\
c^r_i(x_1,\cdots, x_n)= c^r_i(r_1^{-1}(x_1),\cdots, r_n^{-1}(x_n)).
\end{array}
\end{align}
\item $G$ is renaming symmetric if there exists a renaming $r=(r_1,\cdots,r_n)$ such that $G^r$ is ordinary symmetric.
\end{enumerate}
\end{dfn}

\begin{exa}\label{e3.2.2} Consider the Battle of the Sexes, where Player $1$ is husband, Player $2$ is wife, $F$ is football and $C$ is concert. The payoff bi-matrix is as in Table \ref{Tab3.2.1}

\begin{table}[!htbp] %Default value is [tbp]
\centering \caption{Payoff Bi-matrix of Battle of the Sexes\label{Tab3.2.1}}
%\vskip 2mm
\doublerulesep 0.5pt
\begin{tabular}{|c||c|c|}
\hline
\hline $P_1 \backslash P_2$&$F$&$C$\\
\hline $F$& $2,~1$&$0,~0$\\
\hline $C$&$0,~0$&$1,~2$ \\
\hline
\end{tabular}
\end{table}
It is obvious that this game is not ordinary symmetric. But if we rename the strategies in $S_2$ as $F\sim 2$ and $C\sim 1$, i.e., choose $r_1=\id$ and $r_2=M_n$, then it becomes ordinary symmetric. Hence Battle of the Sexes is renaming symmetric.
\end{exa}

Similar to Proposition \ref{p2.4.2}, we can have the following result.

%\begin{prp}\label{p3.2.3}  $G\in {\cal G}_{[n;\kappa]}$ is renaming symmetric with rename $r=(r_1,\cdots,r_n)$, if and only if, for any $\sigma\in {\bf S}_n$
%\begin{align}\label{3.2.1}
%V^c_i=V^c_{\sigma(i)}T^r_{\sigma},\quad i=1,\cdots,n,
%\end{align}
%where
%\begin{align}\label{3.2.2}
%\begin{array}{ccl}
%T^r_{\sigma}&=&D_{\sigma^{-1}(1)}\Phi_{\pi_{\sigma}^{-1}(1)}*D_{\sigma^{-1}(2)}\Phi_{\pi_{\sigma}^{-1}(2)}\\
%~&~&*\cdots*D_{\sigma^{-1}(n)}\Phi_{\pi_{\sigma}^{-1}(n)},
%\end{array}
%\end{align}
%and
%\begin{align}\label{3.2.3}
%D_i=P_{r_i},\quad i=1,2,\cdots,n.
%\end{align}
%\end{prp}

\begin{prp}\label{p3.2.3}  $G\in {\cal G}_{[n;\kappa]}$ is renaming symmetric with rename $r=(r_1,\cdots,r_n)$, if and only if, for any $\sigma\in {\bf S}_n$
\begin{align}\label{3.2.1}
V^c_i=V^c_{\sigma(i)} T^r_{\sigma},\quad i=1,\cdots,n,
\end{align}
where
\begin{align}\label{3.2.2}
\begin{array}{ccl}
T^r_{\sigma}=\Gamma_r^T T_{\sigma}\Gamma_r
\end{array}
\end{align}
and
\begin{align}\label{3.2.3}
\Gamma_r = P_{r_1}\ot P_{r_2}\ot \cdots \ot P_{r_n}.
\end{align}
\end{prp}
\noindent{\it Proof}.
Denote the structure vectors of the renaming symmetric game as $V^r_i$, $i=1,\cdots,n$.

By definition, we have
$$
\begin{array}{ccl}
V^r_i\ltimes_{j=1}^nx_j&=&V^c_i\ltimes_{j=1}^n r_j^{-1}(x_j)\\
~&=&V^c_i\ltimes_{j=1}^n\left(P_{r_j}^{-1}x_j\right)\\
~&=&V^c_i\Gamma_r^{-1}\ltimes_{j=1}^nx_j\\
~&=&V^c_i\Gamma_r^{T}\ltimes_{j=1}^nx_j,
\end{array}
$$
where $\Gamma_r$ is defined in (\ref{3.2.3}).
Hence we have
\begin{align}\label{3.2.301}
V^r_i=V^c_i\Gamma_r^{T}.
\end{align}
On the other hand, Using the fact that $G^r$ is ordinary symmetric, we have, in light of Proposition \ref{p2.4.2},
\begin{align}\label{3.2.302}
\begin{array}{ccl}
V^r_i&=&V^r_{\sigma(i)}T_{\sigma}=V^c_{\sigma(i)}\Gamma_r^TT_{\sigma}.
\end{array}
\end{align}
Comparing (\ref{3.2.301}) with (\ref{3.2.302}) yields (\ref{3.2.1})-(\ref{3.2.2}).
\qed

The following result is parallel to Theorem \ref{t2.4.3} with a mimic proof.

\begin{thm}\label{t3.2.4} Given a renaming $r=(r_1,r_2,\cdots,r_n)$, where $r_i\in {\bf S}_{\kappa}$.
\begin{enumerate}
\item The mapping $\varphi^r:{\bf S}_n\ra \GL({\cal G}_{[n;\kappa]})$ defined as
\begin{align}\label{3.2.4}
\varphi^r(\sigma):=P_{\sigma}\otimes T^r_{\sigma},\quad \sigma\in {\bf S}_n
\end{align}
is a linear representation.
\item $G\in {\cal G}_{[n;\kappa]}$ is renaming symmetric, if and only if, there exists a renaming such that $V_G$ is invariant with respect to the linear representation $\varphi^r$.
\end{enumerate}
\end{thm}

\begin{rem}\label{r3.2.5} Both weighted symmetry and renaming symmetry are theoretically almost trivial. In fact, any property of ordinary symmetry has its corresponding property for weighted or renaming symmetry with some obvious modifications. But from application point of view, the two generalizations enlarged the region of symmetric games a lot, which tremendously  expands the application of symmetric game theory.
\end{rem}

\subsection{Verification of Symmetry}%s-4.3

Let ${\bf S}$ be a finite group and $V$ be a finite dimensional vector space, and $\varphi:{\bf S}\ra \GL(V)$ be a linear representation. Then $W\subset V$ is called a row (column) invariant subspace of $\varphi$, if
$W\varphi(\sigma)\subset W$,
(correspondingly,~~$\varphi(\sigma)W \subset W$), $\forall \sigma\in {\bf S}$.

In this paper, taking the structure of $V_G$ into consideration, the row invariance is considered as a default invariance.

The following proposition comes from the definition of linear representation immediately.

\begin{prp}\label{p3.3.1}
Let $\varphi: {\bf S}\ra \GL(V)$ be a linear representation, and
$$
\Omega:=\{\omega_1,\omega_2,\cdots,\omega_s\}\subset {\bf S}
$$
be a set of generators of ${\bf S}$, i.e., ${\bf S}=\left<\Omega\right>$.
Then $W\subset V$ is invariant with respect to $\varphi$, if and only if,
\begin{align}\label{3.3.0}
W\varphi(\omega)\subset W,\quad \forall \omega\in \Omega.
\end{align}
\end{prp}

This proposition makes the verification of ordinary (or weighted, or renaming ) symmetry easier. Because according to Corollary \ref{c2.4.6}, Theorem \ref{t3.1.5} or Theorem \ref{t3.2.4},  instead verifying the invariance with respect to all elements of ${\bf S}_n$, we need only to verify it for a set of generators.

Next, we recall some standard sets of generators of ${\bf S}_n$.

\begin{prp}\label{p3.3.2}\cite{jac85}
\begin{enumerate}
\item ${\bf S}_n$ is generated by transpositions (switches). That is,
\begin{align}\label{3.3.1}
{\bf S}_n=\left<(i,j)\;|\; 1\leq i<j\leq n \right>.
\end{align}
\item  ${\bf S}_n$ is generated by transpositions with $1$. That is,
\begin{align}\label{3.3.2}
{\bf S}_n=\left<(1,r)\;|\; 1 < j\leq n \right>.
\end{align}
\end{enumerate}
\end{prp}

\begin{exa}\label{e3.3.3} Recall Example \ref{e2.4.7}. Since
$$
{\bf S}_3=\left<  \sigma_3=(1,2), \sigma_6=(1,3)\right>,
$$
 to check whether a game $G$ is ordinary symmetric, it is enough to check (\ref{2.4.9}), for $\sigma_3$ and $\sigma_6$ only. A numerical verification shows the same result, as shown in (\ref{2.4.10}), can be obtained.
\hfill $\Box$
\end{exa}

\section{Name-irrelevant Symmetry and Its Linear Representation}%s-5

This section considers the strategy permutations. We need some preparations.

\begin{dfn}\label{d4.1} Consider ${\bf S}_{n\kappa}$, which can be considered as a set of permutations on grouped (or double-labeled) set
$$
\begin{array}{ccl}
{\cal O}&:=&\{(1,1),(1,2),\cdot,(1,\kappa);\cdots;(n,1),(n,2),\cdot,(n,\kappa)\}\\
~&:=&\{{\cal O}_1,{\cal O}_2,\cdots,{\cal O}_n\}.
\end{array}
$$
$\theta\in {\bf S}_{n\kappa}$ is called a block permutation if there exists an object $(i,\a)$ such that
$\theta((i,\a))=(j,\b)$, then then $\theta(i,\cdot): {\cal O}_i\ra {\cal O}_j$ is a bijection. The set of block permutations is denoted as ${\bf \Theta}_{[n;\kappa]}$.
\end{dfn}

According to above definition, it is easy to verify the following

\begin{prp}\label{p4.2}
\begin{enumerate}
\item $\theta\in {\bf \Theta}_{[n;\kappa]}$, if and only if, there exist $\pi_{\theta}\in {\bf S}_n$, which is a permutation on $\{{\cal O}_i\;|\;i=1,\cdots,n\}$, and $d^{\theta}_1,\cdots,d^{\theta}_n\in {\bf S}_{\kappa}$, where
    \begin{align}\label{4.0}
    d^{\theta}_i=\theta|_{\pi_{\theta}^{-1}(i)}:{\cal O}_{\pi_{\theta}^{-1}(i)}\ra {\cal O}_i,
    \end{align}
    such that
\begin{align}\label{4.1}
P_{\theta}=D^{\theta} P_{\pi_{\theta}},
\end{align}
where
$$
D^{\theta}=\diag(D^{\theta}_1,D^{\theta}_2,\cdots,D^{\theta}_n),
$$
and $D^{\theta}_i$ is the structure matrix of $d^{\theta}_i$, $i=1,\cdots, n$.
\item ${\bf \Theta}_{[n;\kappa]}<{\bf S}_{n\kappa}$ and $\left|{\bf \Theta}_{[n;\kappa]}\right|=n!(\kappa!)^n$.
\end{enumerate}
\end{prp}

Next, we give an alternative expression of $\theta\in {\bf \Theta}_{[n;\kappa]}$. Denote
\begin{align}\label{4.2}
M_{\theta}:=\left(P_{\pi_{\theta}};D_1^{\theta},\cdots,D_n^{\theta}\right),
\end{align}
where
$$
D_i^{\theta}:=P_{\theta|_{\pi_{\theta}^{-1}(i)}},\quad i=1,\cdots,n.
$$
Denote the set of such $M_{\theta}$ by ${\bf M}_{[n;\kappa]}$. That is
\begin{align}\label{4.3}
{\bf M}_{[n;\kappa]}:=\left\{M_{\theta}\;|\; \theta\in {\bf \Theta}_{[n;\kappa]} \right\}.
\end{align}

Let
$$
\begin{array}{l}
M_{\a}=\left(P_{\pi_{\a}};D^{\a}_1,\cdots,D^{\a}_n\right)\in {\bf M}_{[n;\kappa]},\\ 
M_{\b}=\left(P_{\pi_{\b}};D^{\b}_1,\cdots,D^{\b}_n\right)\in {\bf M}_{[n;\kappa]}.
\end{array}
$$
Define a product on ${\bf M}_{[n;\kappa]}$ as
\begin{align}\label{4.4}
M_{\b\circ \a}:=\left(P_{\pi_{\b\circ \a}};D^{\b\circ \a}_1,\cdots,D^{\b\circ \a}_n\right),
\end{align}
where
\begin{align}\label{4.5}
\begin{array}{l}
P_{\pi_{\b\circ\a}}=P_{\pi_{\b}}P_{\pi_{\a}}\\
D^{\b\circ \a}_i=D^{\b}_{i}D^{\a}_{\pi^{-1}_{\b}(i)},\quad i=1,\cdots,n.
\end{array}
\end{align}
Then it is easy to verify that $\left({\bf M}_{[n;\kappa]},\circ\right)$ is a group.

 Let $\theta\in {\bf \Theta}_{[n;\kappa]}$ and set
 \begin{align}\label{4.6}
 {\bf P}_{[n;\kappa]}:=\left\{P_{\theta}\;|\;\theta\in {\bf \Theta}_{[n;\kappa]}\right\}.
 \end{align}
Then it is ready to verify that
$$
{\bf P}_{[n;\kappa]}<\GL(n\kappa,\R).
$$
Moreover, a straightforward computation shows the following result:

\begin{prp}\label{p4.3} $\varphi:{\bf P}_{[n;\kappa]}\ra {\bf M}_{[n;\kappa]}$ as
\begin{align}\label{4.7}
\varphi(P_{\theta}):=\left(P_{\pi_{\theta}};P_{\theta|_{\pi_{\theta}^{-1}(1)}},\cdots, P_{\theta|_{\pi_{\theta}^{-1}(n)}} \right),\quad \theta\in {\bf \Theta}_{[n;\kappa]}.
\end{align}
is a group isomorphism.
\end{prp}

We use an example to depict this.

\begin{exa}\label{e4.4}
Consider $\a=(1,6,3,4,2,5)\in {\bf S}_9$, and $\b=(1,2,3)(4,9,6,8)(5,7)\in {\bf S}_9$. Assume they are acting on
$$
({\cal O}^1,{\cal O}^2,{\cal O}^3):=({\cal O}^1_1,{\cal O}^1_2,{\cal O}^1_3;{\cal O}^2_1,{\cal O}^2_2,{\cal O}^2_3;{\cal O}^3_1,{\cal O}^3_2,{\cal O}^3_3),
$$ which is a set of blocked elements with $n=3$ and $\kappa=3$.
Then we discuss the matrix expression
\begin{itemize}

\item It is easy to verify that both $\a,~\b\in {\bf \Theta}_{[3;3]}$. Moreover, $\pi_{\a}=(1,2)$ and $\pi_{\b}=(2,3)$.

\item Their structure matrices are
$$
P_{\a}=\begin{bmatrix}
0&0&0&0&1&0&0&0&0\\
0&0&0&1&0&0&0&0&0\\
0&0&0&0&0&1&0&0&0\\
0&0&1&0&0&0&0&0&0\\
0&1&0&0&0&0&0&0&0\\
1&0&0&0&0&0&0&0&0\\
0&0&0&0&0&0&1&0&0\\
0&0&0&0&0&0&0&1&0\\
0&0&0&0&0&0&0&0&1\\
\end{bmatrix};\;
P_{\b}=\begin{bmatrix}
0&0&1&0&0&0&0&0&0\\
1&0&0&0&0&0&0&0&0\\
0&1&0&0&0&0&0&0&0\\
0&0&0&0&0&0&0&1&0\\
0&0&0&0&0&0&1&0&0\\
0&0&0&0&0&0&0&0&1\\
0&0&0&0&1&0&0&0&0\\
0&0&0&0&0&1&0&0&0\\
0&0&0&1&0&0&0&0&0\\
\end{bmatrix}
$$

\item
$$
\begin{array}{l}
\a_1=\a|_{\pi_{\a}^{-1}(1)}={\a}|_{{\cal O}^2}=(1,2)\\
P_{\a_1}=\begin{bmatrix}0&1&0\\1&0&0\\0&0&1\end{bmatrix}\\
\a_2=\a|_{\pi_{\a}^{-1}(2)}={\a}|_{{\cal O}^1}=(1,3)\\
P_{\a_2}=\begin{bmatrix}0&0&1\\0&1&0\\1&0&0\end{bmatrix}\\
\a_3=\a|_{\pi_{\a}^{-1}(3)}={\a}|_{{\cal O}^3}=\id\\
P_{\a_3}=I_3\\
\end{array}
$$
$$
\begin{array}{l}
\b_1=\b|_{\pi_{\b}^{-1}(1)}={\b}|_{{\cal O}^1}=(1,2,3)\\
P_{\b_1}=\begin{bmatrix}0&0&1\\1&0&0\\0&1&0\end{bmatrix}\\
\b_2=\b|_{\pi_{\b}^{-1}(2)}={\b}|_{{\cal O}^3}=(1,2)\\
P_{\b_2}=\begin{bmatrix}0&1&0\\1&0&0\\0&0&1\end{bmatrix}\\
\b_3=\b|_{\pi_{\b}^{-1}(3)}={\b}|_{{\cal O}^2}=(1,3,2)\\
P_{\b_3}=\begin{bmatrix}0&1&0\\0&0&1\\1&0&0\end{bmatrix}\\
\end{array}
$$
\item
Consider $\gamma=\b\circ \a$. It is easy to calculate that
$$
\gamma=(1,8,4,3,9,6)(2,7,5).
$$
$$
\pi_{\gamma}=\pi_{\b}\circ \pi_{\a}=(1,3,2).
$$
\item Since $P_{\a},~P_{\b}\in {\bf P}_{[3;3]}$, we have
$$
P_{\gamma}=P_{\b} P_{\a}=\begin{bmatrix}
 0  & 0  & 0  & 0  & 0  & 1  & 0  & 0  & 0\\
 0  & 0  & 0  & 0  & 1  & 0  & 0  & 0  & 0\\
 0  & 0  & 0  & 1  & 0  & 0  & 0  & 0  & 0\\
 0  & 0  & 0  & 0  & 0  & 0  & 0  & 1  & 0\\
 0  & 0  & 0  & 0  & 0  & 0  & 1  & 0  & 0\\
 0  & 0  & 0  & 0  & 0  & 0  & 0  & 0  & 1\\
 0  & 1  & 0  & 0  & 0  & 0  & 0  & 0  & 0\\
 1  & 0  & 0  & 0  & 0  & 0  & 0  & 0  & 0\\
 0  & 0  & 1  & 0  & 0  & 0  & 0  & 0  & 0\\
\end{bmatrix}\in {\bf P}_{[3;3]}.
$$
It follows that
\begin{align}\label{4.8}
\begin{array}{l}
\gamma_1=\gamma|_{\pi_{\gamma}^{-1}(1)}={\gamma}|_{{\cal O}^2}=(1,3)\\
P_{\gamma_1}=\begin{bmatrix}0&0&1\\0&1&0\\1&0&0\end{bmatrix}\\
\gamma_2={\gamma}|_{\pi_{\gamma}^{-1}(2)}={\gamma}|_{{\cal O}^3}=(1,2)\\
P_{\gamma_2}=\begin{bmatrix}0&1&0\\1&0&0\\0&0&1\end{bmatrix}\\
{\gamma}_3={\gamma}|_{\pi_{\gamma}^{-1}(3)}={\gamma}|_{{\cal O}^1}=(1,2)\\
P_{\gamma_3}=\begin{bmatrix}0&1&0\\1&0&0\\0&0&1\end{bmatrix}\\
\end{array}
\end{align}
\item We already know that ${\bf \Theta}_{[n;\kappa]}$ is isomorphic to ${\bf P}_{[n;\kappa]}$. To check that ${\bf M}_{[n;\kappa]}$ is also isomorphic to them, we use formula (\ref{4.6}) to calculate $D^{\gamma}_1$, $D^{\gamma}_2$, and $D^{\gamma}_3$:
$$
\begin{array}{ccl}
D^{\gamma}_1&=&P_{{\gamma}_1}=P_{\b|_{\pi_{\b}^{-1}(1)}}P_{\a|_{\pi_{\a}^{-1}(\pi_{\b}^{-1}(1))}}\\
~&=&P_{{\b}|_{{\cal O}^1}}P_{{\a}|_{\pi_{\a}^{-1}(1)}}\\
~&=&P_{\b|_{{\cal O}^1}}P_{{\a}|_{{\cal O}^1}}\\
~&=&\begin{bmatrix}0&0&1\\1&0&0\\0&1&0\end{bmatrix}\begin{bmatrix}0&1&0\\1&0&0\\0&0&1\end{bmatrix}
=\begin{bmatrix}0&0&1\\0&1&0\\1&0&0\end{bmatrix};\\
\end{array}
$$
$$
\begin{array}{ccl}
D^{\gamma}_2&=&P_{{\gamma}_2}=P_{\b|_{\pi_{\b}^{-1}(2)}}P_{{\a}|_{\pi_{\a}^{-1}(\pi_{\b}^{-1}(2))}}\\
~&=&P_{{\b}|_{{\cal O}^2}}P_{\a|_{\pi_{\a}^{-1}(3)}}\\
~&=&P_{\b|_{{\cal O}^2}}P_{\a|_{{\cal O}^3}}\\
~&=&\begin{bmatrix}0&1&0\\1&0&0\\0&0&1\end{bmatrix}\begin{bmatrix}1&0&0\\0&1&0\\0&0&1\end{bmatrix}
=\begin{bmatrix}0&1&0\\1&0&0\\0&0&1\end{bmatrix};\\
\end{array}
$$
$$
\begin{array}{ccl}
D^{\gamma}_3&=&P_{{\gamma}_3}=P_{\b|_{\pi_{\b}^{-1}(3)}}P_{\a|_{\pi_{\a}^{-1}(\pi_{\b}^{-1}(3))}}\\
~&=&P_{{\b}|_{{\cal O}^3}}P_{\a|_{\pi_{\a}^{-1}(2)}}\\
~&=&P_{\b|_{{\cal O}^3}}P_{\a|_{x^2}}\\
~&=&\begin{bmatrix}0&1&0\\0&0&1\\1&0&0\end{bmatrix}\begin{bmatrix}0&0&1\\0&1&0\\1&0&0\end{bmatrix}
=\begin{bmatrix}0&1&0\\1&0&0\\0&0&1\end{bmatrix}.\\
\end{array}
$$
These results coincide with the results in (\ref{4.8}).
\end{itemize}
\end{exa}

\begin{dfn}\label{d4.5} A normal strategy permutation of $G\in {\cal G}_{[n;\kappa]}$ is a permutation $\theta\in {\bf S}_{n\kappa}$, satisfying that for each $i$ there exists $j=\pi(i)$, such that $\theta(S_i)= S_j$, and hence $\theta\big|_{S_i}\in {\bf S}_{\kappa}$, $i=1,\cdots,n$.
\end{dfn}

Note that since $\theta\in {\bf S}_{n\kappa}$, it is forced that $\pi\in {\bf S}_n$. To emphasize that $\pi$ is induced from $\theta$, it is denoted as $\pi_{\theta}$.

Now, it is clear that a normal strategy permutation is a block permutation. That is, $\theta\in {\bf \Theta}_{[n;\kappa]}$.

\begin{dfn}\label{d4.6} \cite{nas51} Consider a $G\in {\cal G}_{[n;\kappa]}$, a normal strategy permutation $\theta\in {\bf \Theta}_{[n;\kappa]}$ is called a strategy symmetry of game $G$ if
\begin{align}\label{4.9}
\begin{array}{l}
c_i(x_1,\cdots,x_n)=c_{\pi_{\theta}(i)}\left(d_1^{\theta}(x_{\pi_{\theta}^{-1}(1)}),\cdots, d_n^{\theta}(x_{\pi_{\theta}^{-1}(n)})\right),\\
i=1,\cdots,n,
\end{array}
\end{align}
where
\[
d_i^{\theta} := \theta|_{\pi_{\theta}^{-1}(i)}: S_{\pi_{\theta}^{-1}(i)}\ra S_i.
\]
The set of strategy symmetries of $G$ is denoted as ${\bf \Theta}(G)$.
\end{dfn}

Note that $\theta|_{\pi_{\theta}^{-1}(i)}: S_{\pi_{\theta}^{-1}(i)}\ra S_i$, which can be expressed as follows: Assume $\d_{\kappa}^{\a}\in  S_{\pi_{\theta}^{-1}(i)}$, then $\theta|_{\pi_{\theta}^{-1}(i)}: \d_{\kappa}^{\a}\mapsto \d_{\kappa}^{\b}$, which is simply considered as a mapping: $\a\mapsto \b$. From this point of view,  $\theta|_{\pi_{\theta}^{-1}(i)}\in {\bf S}_{\kappa}$, and its structure matrix is denoted as
$$
P_{\theta|_{\pi_{\theta}^{-1}(i)}}:=D_i^{\theta}\in {\bf P}_{\kappa},\quad i=1,\cdots,n.
$$
Using Proposition \ref{p2.3.2} and the expression (\ref{2.1.4}), a straightforward computation shows that (\ref{4.9}) can be expressed as
\begin{align}\label{4.10}
\begin{array}{ccl}
V^c_ix&=&V^c_{\pi_{\theta}(i)}\ltimes_{j=1}^n D^{\theta}_jx_{\pi_{\theta}^{-1}(j)}\\
~&=&V^c_{\pi_{\theta}(i)}\ltimes_{j=1}^n D^{\theta}_j\Phi_{\pi_{\theta}^{-1}(j)}x\\
~&=&V^c_{\pi_{\theta}(i)}\left(D_1^{\theta}\ot\cdots\ot D_n^{\theta}\right)T_{\pi_{\theta}}x\\
~&:=&V^c_{\pi_{\theta}(i)}T^{\theta}x.
\end{array}
\end{align}

Summarizing the above argument yields
\begin{prp}\label{p4.7}  Consider a $G\in {\cal G}_{[n;\kappa]}$, and a normal strategy permutation $\theta\in {\bf \Theta}_{[n;\kappa]}$. $\theta$ is a strategy symmetry of game $G$, if and only if,
\begin{align}\label{4.11}
V^c_i=V^c_{\pi_{\theta}(i)}T^{\theta},\quad i=1,\cdots,n,
\end{align}
where
\begin{align}\label{4.12}
T^{\theta}=\left(D_1^{\theta}\ot\cdots\ot D_n^{\theta}\right)\left(\Phi_{\pi_{\theta}^{-1}(1)}*\cdots* \Phi_{\pi_{\theta}^{-1}(n)} \right).
\end{align}
\end{prp}

The following proposition is a modification of the corresponding result from \cite{cao16}.

\begin{prp}\label{p4.8} Denote the set of strategy symmetries of $G$ by $\Theta(G)$. Then
\begin{align}\label{4.13}
{\bf \Theta}(G)<{\bf \Theta}_{[n;\kappa]}<{\bf S}_{n\kappa}.
\end{align}
\end{prp}
Therefore, ${\bf \Theta}(G)$ is called the strategy symmetric group of $G$.

Similar to Corollary \ref{c2.4.6}, Theorem \ref{t3.1.5} and Theorem \ref{t3.2.4}, by taking Propositions \ref{p3.3.1}-\ref{p3.3.2} into consideration, we have the following result:

\begin{thm}\label{t4.801} Consider a $G\in {\cal G}_{[n;\kappa]}$. Assume $\theta\in {\bf \Theta}_{[n;\kappa]}$.
\begin{enumerate}
\item $\varphi^n:\Theta_{[n;\kappa]}\ra {\cal G}_{[n;\kappa]}$ defined by
\begin{align}\label{4.1301}
\varphi^n(\theta)=P_{\pi_{\theta}}\otimes T^{\theta}\in \GL({\cal G}_{[n;\kappa]})
\end{align}
is a linear representation.
\item $\theta\in {\bf \Theta}_{[n;\kappa]}$ is a normal strategy permutation of $G$, i.e., $\theta\in {\bf \Theta}(G)$, if and only if, $V_G$ is invariant with respect to $\varphi^n(\theta)$.
\end{enumerate}
\end{thm}

\begin{exa}\label{e4.9} Given $G\in {\cal G}_{[3;3]}$ and $\theta=\b\in {\bf \Theta}_{[3;3]}$, where $\b$ is as in Example \ref{e4.4}. The question is: when $\theta\in {\bf \Theta}(G)$ ?

First, we calculate $T_{\theta}$ as
$$
\begin{array}{ccl}
T^n_{\theta}&=&\left(D_1^{\theta}\ot D_2^{\theta} \ot \cdots\ot D_n^{\theta}\right)T_{\pi_{\theta}}\\
~&=&\d_{27}[15,12,18,13,10,16,14,11,17,24,21,27,22,\\
~&~&~~~~~19,25,23,20,26,6,3,9,4,1,7,5,2,8].
\end{array}
$$
According to Proposition \ref{p4.7}, $\theta\in {\bf \Theta}(G)$, if and only if
\begin{align}\label{4.14}
\begin{cases}
V^c_1=V^c_1T^{\theta}\\
V^c_2=V^c_3T^{\theta}\\
V^c_3=V^c_2T^{\theta}\\
\end{cases}
\end{align}
(\ref{4.14}) is equivalent to
\begin{align}\label{4.15}
\begin{cases}
V^c_1=V^c_1T^{\theta}\\
V^c_2=V^c_2(T^{\theta})^2\\
V^c_3=V^c_2T^{\theta}\\
\end{cases}
\end{align}
It follows that the necessary and sufficient condition is:
\begin{align}\label{4.16}
\begin{array}{ccl}
V^c_1&=&[a,b,b,c,a,a,a,b,b,a,b,b,c,a,a,a,\\
~&~&~~~b,b,a,b,b,c,a,a,a,b,b];\\
V^c_2&=&[d,e,g,f,g,g,d,g,e,d,e,g,f,g,g,d,\\
~&~&~~~g,e,d,e,g,f,g,g,d,g,e];\\
V^c_3&=&[g,g,e,f,d,d,g,e,g,g,g,e,f,d,d,g,\\
~&~&~~~e,g,g,g,e,f,d,d,g,e,g],\\
\end{array}
\end{align}
where $a,b,c,d,e,f,g$ are arbitrary real numbers.

\end{exa}

\begin{dfn}\label{d4.10} \cite{cao16} $\pi\in {\bf S}_n$ is called a name-irrelevant player symmetry if there exists a strategy symmetry $\theta$ such that $\pi=\pi_{\theta}$. The set of name-irrelevant player symmetries is denoted by $\Pi(G)$. That is,
\begin{align}\label{4.17}
\Pi(G)=\left\{\pi_{\theta}\;|\;\theta\in \Theta(G)\right\}.
\end{align}
\end{dfn}

It follows from the definition that $\Pi(G)\subset {\bf S}_n$ is a subgroup.

\begin{dfn}\label{d4.11} \cite{pel99}
$G\in {\cal G}_{[n;\kappa]}$ is said to be name-irrelevant symmetric, if every player permutation is a name-irrelevant player symmetry, i.e., $\Pi(G)={\bf S}_n$.
\end{dfn}

The following example gives the same conclusion as in \cite{cao16} with detailed computation.

\begin{exa}\label{e4.12}  Matching Pennies has payoff bi-matrix as in Table \ref{Tab2.4.1}.

\begin{table}[!htbp] %Default value is [tbp]
\centering \caption{Payoff Bi-matrix of Matching Pennies}\label{Tab2.4.1}
%\vskip 2mm
\doublerulesep 0.5pt
\begin{tabular}{|c||c|c|}
\hline
\hline $P_1 \backslash P_2$&$1$&$2$\\
\hline $1$& $1,~-1$&$-1,~1$\\
\hline $2$&$-1,~1$&$1,~-1$ \\
\hline
\end{tabular}
\end{table}

It follows from Table \ref{Tab2.4.1} that
$$
V^c_1=[1,-1,-1,1];\quad V^c_2=[-1,1,1,-1].
$$
Denote
$$
{\bf \Theta}_{[2;2]}=\{\theta_i\;|\; i=1,2,3,4,5,6,7,8\},
$$
where
$$
\begin{array}{ll}
P_{\theta_1}=\begin{bmatrix}I_2&0\\0&I_2\end{bmatrix};&
P_{\theta_2}=\begin{bmatrix}I_2&0\\0&M_n\end{bmatrix}\\
P_{\theta_3}=\begin{bmatrix}M_n&0\\0&I_2\end{bmatrix};&
P_{\theta_4}=\begin{bmatrix}M_n&0\\0&M_n\end{bmatrix}\\
P_{\theta_5}=\begin{bmatrix}0&I_2\\I_2&0\end{bmatrix};&
P_{\theta_6}=\begin{bmatrix}0&I_2\\M_n&0\end{bmatrix}\\
P_{\theta_7}=\begin{bmatrix}0&M_n\\I_2&0\end{bmatrix};&
P_{\theta_8}=\begin{bmatrix}0&M_n\\M_n&0\end{bmatrix}.\\
\end{array}
$$
Then,
$$
\pi_{\theta_i}=
\begin{cases}
\id,\quad i=1,2,3,4,\\
(1,2),\quad i=5,6,7,8.
\end{cases}
$$
Using (\ref{2.4.8}), we can calculate that
$$
T^{\theta_i}=
\begin{cases}
\left(D^{\theta_i}_1\ot D^{\theta_i}_2\right)\left((I_2\otimes {\bf 1}_2^T)*({\bf 1}_2^T\otimes I_2)\right)\; i=1,2,3,4,\\
\left(D^{\theta_i}_1\ot D^{\theta_i}_2\right)\left(({\bf 1}_2^T\otimes I_2)*(I_2\otimes {\bf 1}_2^T)\right)\; i=5,6,7,8.\\
\end{cases}
$$
According to Proposition \ref{p4.7}, $\theta_i\in {\bf \Theta}(G)$, if and only if
$$
\begin{cases}
V^c_1=V^c_1T^{\theta_i},\\
V^c_2=V^c_2T^{\theta_i},\quad i=1,2,3,4,\\
\end{cases}
$$
or
$$
\begin{cases}
V^c_1=V^c_2T^{\theta_i},\\
V^c_2=V^c_1T^{\theta_i},\quad i=5,6,7,8.\\
\end{cases}
$$
A straightforward verification shows that
$$
{\bf \Theta}(G)=\left\{\theta_1,\theta_4,\theta_6,\theta_7\right\}.
$$
It is obvious that $\Pi(G)={\bf S}_2$. Hence, $G$ is of name-irrelevant player symmetry.
\end{exa}

%
%\begin{exa}\label{e2.4.11} Given $G\in {\cal G}_{[3;3]}$ and $\theta\in {\bf \Theta}_{[3;3]}$. Assume $\theta =\a$, where $\a$ is as in Example \ref{e2.4.402}. Similar to Example \ref{e2.4.801}, we can prove that $\a\in {\bf \Theta}(G)$, if and only if,
%\begin{align}\label{2.4.13}
%\begin{array}{ccl}
%V^c_1&=&[a,b,c,d,e,f,g,h,i,d,e,f,g,h,i,\\
%~&~&~~~a,b,c,g,h,i,a,b,c,d,e,f];\\
%V^c_2&=&[a,b,c,g,h,i,d,e,f,g,h,i,d,e,f,\\
%~&~&~~~a,b,c,d,e,f,a,b,c,g,h,i];\\
%V^c_3&=&[A,B,C,D,E,F,D,E,F,D,E,F,D,E,F,\\
%~&~&~~~A,B,C,D,E,F,A,B,C,D,E,F],\\
%\end{array}
%\end{align}
%where $a,b,c,d,e,f,g,h,i,A,B,C,D,E,F$ are arbitrary real numbers.
%
%Since $\left<\a,~\b\right> = {\bf S}_3$, where $\left<\a,\b\right>$ is the group generated by $\a$ and $\b$, if $\a,~\b\in {\bf \Theta}(G)$, then $G$ is name-irrelevant symmetric, that is, $\Pi(G)={\bf S}_3$. Unfortunately, if $V^c_i$, $i=1,2,3$ satisfies
%both (\ref{2.4.11}) and (\ref{2.4.13}), then $V^c_1=V^c_2=V^c_3$ are constant. The game is meaningless.
%
%\end{exa}

\section{Potential Game}%s-6

\begin{dfn}\label{d5.1.1} Let $G=(N,S,C)\in {\cal G}_{[n;k_1,\cdots,k_n]}$. $G$ is called a weighted potential game, if there exist a set of weights $w_i>0$, $i=1,\cdots,n$, and a function $P:S\ra \R$, called the potential function, such that
\begin{align}\label{5.1.1}
\begin{array}{l}
c_i(x_i,s_{-i})-c_i(y_i,s_{-i})=w_i[ P(x_i,s_{-i})-P(y_i,s_{-i})],\\
~~~ x_i,y_i\in S_i, s_{-i}\in S_{-i},\; i=1,\cdots,n.
\end{array}
\end{align}
When $w_i=1$, $i=1,\cdots,n$, $G$ is called a (exact) potential game.
\end{dfn}

Denote
\begin{align}\label{5.1.2}
k^{[i,j]}=\begin{cases}
1,\quad i>j\\
\prod_{s=i}^jk_s,\quad 1\leq i\leq j\leq n.
\end{cases}
\end{align}
Using (\ref{5.1.2}), we define
\begin{align}\label{5.1.3}
E_i:=I_{k^{[1,i-1]}}\otimes {\bf 1}_{k_i}\otimes I_{k^{[i+1,n]}},\quad i=1,\cdots,n.
\end{align}
Then we construct $E^w(n)$, $\xi^w(n)$, and $b^w(n)$ as follows:

\begin{align}\label{5.1.4}
E^w(n):=\begin{bmatrix}
-w_2E_1&w_1E_2&0&\cdots&0\\
-w_3E_1&0&w_1E_3&\cdots&0\\
\vdots&~&~&~&~\\
-w_nE_1&0&0&\cdots&w_1E_n
\end{bmatrix}
\end{align}

\begin{align}\label{5.1.5}
\xi^w(n)=\begin{bmatrix}
\xi^w_1\\
\xi^w_2\\
\vdots\\
\xi^w_n
\end{bmatrix},
\end{align}
where $\xi_i^w\in \R^{k/k_i}$ ($k=\prod_{s=1}^nk_i$) are unknowns.

\begin{align}\label{5.1.6}
b^w(n)=\begin{bmatrix}
b^w_2\\
b^w_3\\
\vdots\\
b^w_n
\end{bmatrix}=\begin{bmatrix}
\left(w_1 V^c_2-w_2V^c_1\right)^T\\
\left(w_1 V^c_3-w_3V^c_1\right)^T\\
\vdots\\
\left(w_1 V^c_n-w_nV^c_1\right)^T
\end{bmatrix},
\end{align}
where $V^c_i$ is the structure vector of $c_i$, that is
$$
c_i(x_1,\cdots,x_n)=V^c_i\ltimes_{j=1}^nx_j,\quad x_j\in S_j,\;i=1,\cdots,n.
$$
Finally, we build a linear system, called the potential equation, as
\begin{align}\label{5.1.7}
E^w(n)\xi^w(n)=b^w(n).
\end{align}
Similar to the corresponding result in \cite{che14}, we can prove the following result:

\begin{thm}\label{t5.1.2}
A game $G\in {\cal G}_{[n;k_1,\cdots,k_n]}$ is a weighted potential game with $w_i>0$, $i=1,\cdots,n$, if and only if, the equation (\ref{5.1.7}) has solution. Moreover, if a solution exists, then a structure vector of the potential function $P$ is
\begin{align}\label{5.1.8}
V^P=\frac{1}{w_1}\left(V^c_1-\left(\xi^w_1\right)^T E_1^T\right).
\end{align}
\end{thm}

\begin{rem}\label{r5.1.3} When $w_i=1$, $i=1,\cdots,n$, the results become the corresponding results for pure potential games. In this case, the superscripts $w$ in above formulas are all removed.
\end{rem}

%Next, for convenience in the sequel applications we give an alternative form of (\ref{5.1.4})-(\ref{5.1.6}), and for notational ease, set $w_i=1$, $i=1,\cdots,n$.

%\begin{align}\label{5.1.9}
%E(n):=\begin{bmatrix}
%E_1 & -E_2 & & & \\
% & E_2& -E_3& & \\
% & & \ddots & \ddots&\\
% &  & &  E_{n-1}& -E_n
%\end{bmatrix}
%\end{align}
%
%\begin{align}\label{5.1.10}
%\xi(n)=\begin{bmatrix}
%\xi_1\\
%\xi_2\\
%\vdots\\
%\xi_n
%\end{bmatrix},
%\end{align}
%where $\xi_i\in \R^{k/k_i}$ ($k=\prod_{s=1}^nk_i$) are unknowns.
%
%\begin{align}\label{5.1.11}
%b(n)=\begin{bmatrix}
%b_2\\
%b_3\\
%\vdots\\
%b_n
%\end{bmatrix}=\begin{bmatrix}
%\left(V^c_1-V^c_2\right)^T\\
%\left(V^c_2-V^c_3\right)^T\\
%\vdots\\
%\left(V^c_{n-1}-V^c_n\right)^T
%\end{bmatrix},
%\end{align}

%\begin{cor}\label{c5.1.4} A game $G\in {\cal G}_{[n;k_1,\cdots,k_n]}$ is a (pure) potential game, if and only if,
%\begin{align}\label{5.1.12}
%E(n)\xi(n)=b(n)
%\end{align}
%has solution. Moreover, if a solution exists, then a structure vector of the potential function $P$ is
%\begin{align}\label{5.1.13}
%V^P=V^c_1-\xi_1^T E_1^T.
%\end{align}
%\end{cor}

\section{Symmetric Boolean Games}%s-7

\subsection{Matrix Expression of Symmetric Boolean Games}%s-7.1

Assume $G\in {\cal G}_{[n;2]}$, then $G$ is called a Boolean game, because the strategy sets $S_i=\{0,1\}$, $i=1,\cdots,n$. To investigate the ordinary symmetry of Boolean games, we first give an alternative definition. To this end, we introduce a new concept first.

\begin{dfn}\label{d7.1.1} \cite{cao16} Consider a finite game $G=(N,S,C)\in {\cal G}_{[n;\kappa]}$, and let $s\in S$, the strategy multiplicity vector of $s$ is defined as
\begin{align}\label{7.1.1}
\#(s)=\left(\#(s,1),\#(s,2),\cdots, \#(s,\kappa)\right),
\end{align}
where
$$
\#(s,i):=\left|\{s_j\;|\;s_j=i\}\right|,\quad i=1,\cdots,\kappa.
$$
\end{dfn}

Note that if $t\in S^{N_1}:=\prod_{j \in N_1} S_j$, then $\#(t)$ is defined in a similar way. Particularly, $\#(s_{-i})\in \R^{n}$ is well defined.

It is easy to verify the following equivalent condition of ordinary symmetry.

\begin{prp}\label{p7.1.2} Let $G\in {\cal G}_{[n;\kappa]}$ be a finite game. $G$ is weighted ordinary symmetric with respect to $\mu_i$, $i=1,\cdots,n$, if and only if,
\begin{align}\label{7.1.2}
\mu_ic_i(x_i,x_{-i})=\mu_jc_j(y_j,y_{-j}),\quad 1\leq i,j\leq n,
\end{align}
where $x_i=y_j$, and $\#(x_{-i})=\#(y_{-j})$.
\end{prp}
\begin{rem}\label{r7.1.3} When $\mu_i=1$, $i=1,\cdots,n$, (\ref{7.1.2}) becomes a necessary and sufficient condition for ordinary symmetry. It is used as the definition of ordinary symmetry in \cite{cao16}.
\end{rem}

According to (\ref{3.1.1}) with $w_i=1,~\forall i$, we know that a game $G\in {\cal G}_{[n;\kappa]}$ is ordinary symmetric, if and only if, for any $\sigma\in {\bf S}_n$ we have
\begin{align}\label{7.1.3}
V^c_i\ltimes_{i=1}^nx_i=V^c_{\sigma(i)}\ltimes_{i=1}^nx_{\sigma^{-1}(i)}.
\end{align}

Express $c_i$ into algebraic form as
\begin{align}\label{7.1.4}
c_i(x)=V^c_i\ltimes_{j=1}^nx_j=V^c_iW_{[\kappa,\kappa^{i-1}]}x_i\ltimes_{j\neq i}x_{j}.
\end{align}
According to Proposition \ref{p3.3.2} (precisely, equation (\ref{3.3.2}), it is enough to check (\ref{7.1.4}) for $\sigma=(1,s)\in {\bf S}_{n-1}$ ($2\leq s\leq n-1$) on $\{x_j\;|\; j\neq i\}$ will not change $c_i(x)$. We, therefore, have
\begin{align}\label{7.1.5}
\begin{array}{ccl}
c_i(x)&=&V^c_iW_{[\kappa,\kappa^{i-1}]}x_iW_{[\kappa^{s-2},\kappa]}W_{[\kappa,\kappa^{s-1}]}\ltimes_{j\neq i}x_{j}\\
~&=&V^c_iW_{[\kappa,\kappa^{i-1}]}\left(I_{\kappa}\otimes W_{[\kappa^{s-2},\kappa]}W_{[\kappa,\kappa^{s-1}]}\right)x_i\ltimes_{j\neq i}x_{j}\\
~&=&V^c_iW_{[\kappa,\kappa^{i-1}]}\left(I_{\kappa}\otimes W_{[\kappa^{s-2},\kappa]}W_{[\kappa,\kappa^{s-1}]}\right)\\
~&~&~~W_{[\kappa^{i-1},\kappa]}\ltimes_{j=1}^nx_{j}\\
\end{array}
\end{align}
Comparing (\ref{7.1.4}) with (\ref{7.1.5}) yields that
\begin{align}\label{7.1.6}
\begin{array}{l}
V^c_i\left[ W_{[\kappa,\kappa^{i-1}]}\left(I_{\kappa}\otimes W_{[\kappa^{s-2},\kappa]}W_{[\kappa,\kappa^{s-1}]}\right)W_{[\kappa^{i-1},\kappa]}-I_r\right]\\
~~=0,\quad s=2,3,\cdots,n-1,
\end{array}
\end{align}
where $r=max\{\kappa^i,\kappa^{s+1}\}$.

According to Proposition \ref{p3.3.2}, as long as $V^c_i$ satisfies (\ref{7.1.3}), for any $\sigma\in {\bf S}_n$ and $\sigma(i)=i$, we have
\begin{align}\label{7.1.7}
c_i(x)=c_i\left(x_{\sigma^{-1}(1)}, x_{\sigma^{-1}(2)},\cdots, x_{\sigma^{-1}(n)}\right).
\end{align}
Hence, (\ref{7.1.7}) is the necessary and sufficient condition for a single payoff function to be satisfied in a symmetric game.

Next, we consider the condition for cross payoffs, that is, for any two payoff functions in a symmetric game need to be satisfied.
Note that for $p\neq q$, 
\begin{align}\label{7.1.8}
\begin{cases}
c_p(x)=V^c_pW_{[\kappa,\kappa^{p-1}]}x_p\ltimes_{j\neq p}x_j\\
c_q(x)=V^c_qW_{[\kappa,\kappa^{q-1}]}x_q\ltimes_{j\neq q}x_j.
\end{cases}
\end{align}
Assume $x_p=x_q$, then $\#(x_{-p})=\#(x_{-q})$. According to Proposition \ref{p7.1.2}, we have
\begin{align}\label{7.1.9}
V^c_pW_{[\kappa,\kappa^{p-1}]}= V^c_qW_{[\kappa,\kappa^{q-1}]},\quad p\neq q.
\end{align}

Taking $i=1$ for (\ref{7.1.7}) and $p=1$ for(\ref{7.1.9}), it is ready to verify the following result:

\begin{thm}\label{t7.1.4} $G\in {\cal G}_{[n;\kappa]}$ is a symmetric game, if and only if,
\begin{enumerate}
\item[(i)]
\begin{align}\label{7.1.10}
\begin{array}{l}
V^c_1\left[I_{\kappa}\otimes \left(W_{[\kappa^{s-2}, \kappa]}W_{[\kappa, \kappa^{s-1}]}\right) -I_{k^{s+1}}\right]=0,\\
\quad s=2,3,\cdots,n-1.
\end{array}
\end{align}
\item[(ii)]
\begin{align}\label{7.1.11}
V^c_i=V^c_1W_{[\kappa^{i-1}, \kappa]},\quad i=2,3,\cdots,n.
\end{align}
\end{enumerate}
\end{thm}

\subsection{Verification of Symmetric Boolean Games}%s-7.2

This subsection considers when a Boolean game is symmetric. Let $G\in {\cal G}_{[n;2]}$, and according to (\ref{7.1.10}),
\begin{align}\label{7.2.1}
\begin{array}{ccl}
V^c_1&\in& I_2\otimes \left[W_{[2^{s-2},2]} W_{[2, 2^{s-1}]}-I_{2^s} \right]^{\perp}\otimes I_{2^{n-s-1}},\\
~&~&\quad s=2,3,\cdots,n-1.
\end{array}
\end{align}
Using formula (\ref{2.2.9}), we can calculate that
$$
\begin{array}{l}
\left[W_{[2^{s-2},2]} W_{[2, 2^{s-1}]}-I_{2^s} \right]^{\perp}\\
=\left[\left(W_{[2^{s-2},2]}\otimes I_2\right)\left( I_{2^{s-2}}\otimes W_{[2, 2]}\right)\left( W_{[2, 2^{s-2}]}\otimes I_2 \right)\right.\\
~~~\left.-I_{2^s}\right]^{\perp}\\
=\left\{\left(W_{[2^{s-2},2]}\otimes I_2\right)\left[ I_{2^{s-2}}\otimes \left(W_{[2, 2]}-I_4\right)\right]\right.\\ ~~~\left.\left(W_{[2, 2^{s-2}]}\otimes I_2 \right)\right\}^{\perp}\\
=\left(W_{[2^{s-2},2]}\otimes I_2\right)\left[ I_{2^{s-2}}\otimes \left(W_{[2, 2]}+I_4\right)\right]\\
~~~\left(W_{[2, 2^{s-2}]}\otimes I_2 \right)\\
=\left[W_{[2^{s-2},2]} W_{[2, 2^{s-1}]}+I_{2^s} \right].
\end{array}
$$
The penultimate equality comes from (i) the orthogonality and (ii) the dimension complementary.

Next, we investigate the vector space structure of $V^c_1$, satisfying (\ref{7.1.10}). Note that for each $s_{-1}\in S_{-1}$ the $\#(s_{-1},0)$ (or $\#(s_{-1},1)$) could be $\{0,1,2,\cdots,n-1\}$. According to Proposition \ref{p7.1.2},
$$
S_{-1}^j:=\left\{s_{-1}\in S_{-1}\;|\;\#(s_{-1},0)=j\right\},\quad j=0,1,\cdots,n-1
$$
has the same value of $c_1(x_1,x_{-1})$. Hence, for symmetric Boolean games $V^c_1$ is of dimension $2n$, where the $2$ comes from the fact that $x_1$ has two possible choices. But since $V^c_1\in \R^{2^n}$ we construct a mapping $h_n:\R^{2n}\ra \R^{2^n}$, such that a symmetric game satisfies $V^c_1\in Im(h_n)$, where $Im(h_n)$ is the image set of $h_n$.

To construct this $h_n$, we first set
$T_2:=I_2$. Then we inductively construct
\begin{align}\label{7.2.2}
T_{k+1}:=\begin{bmatrix}
T_k&{\bf 0}_{2^{k-1}}\\
{\bf 0}_{2^{k-1}}&T_k
\end{bmatrix},\quad k=2,3,\cdots.
\end{align}
Then we define
\begin{align}\label{7.2.3}
H_k:=I_2\otimes T_k, \quad k=2,3,\cdots.
\end{align}
A straightforward computation shows the following:

\begin{prp}\label{p7.2.1} Let $h_n:\R^{2n}\ra \R^{2^n}$ be the mapping determined by
$$
h_n(x):=H_nx,\quad x\in \R^{2n}.
$$
Then  $V^c_1$ is a suitable candidate of $V_1^c$ for a symmetric Boolean game, if and only if, $V^c_1=h_n^T(x)$, for some $x\in \R^{2n}$.  In other words, for all $x\in \R^{2n}$ that $h_n^T(x)$ satisfy (\ref{7.1.10}), and vise versa.
\end{prp}
\noindent{\it Proof.}  See Appendix A. \hfill $\Box$

\begin{exa}\label{e7.2.2}
\begin{enumerate}
\item Constructing $H_n$:

Since $T_2=I_2$, we have
$$
H_2=I_2\otimes I_2=I_4.
$$
$$
T_3=
\begin{bmatrix}
T_2&{\bf 0}_2\\
{\bf 0}_2&T_2
\end{bmatrix}=\begin{bmatrix}
1&0&0\\
0&1&0\\
0&1&0\\
0&0&1
\end{bmatrix},
$$
$$
T_4=
\begin{bmatrix}
T_3&{\bf 0}_4\\
{\bf 0}_4&T_3
\end{bmatrix}=\begin{bmatrix}
1&0&0&0\\
0&1&0&0\\
0&1&0&0\\
0&0&1&0\\
0&1&0&0\\
0&0&1&0\\
0&0&1&0\\
0&0&0&1\\
\end{bmatrix},
$$
and so on. Then $H_3=I_2\otimes T_3$,  $H_4=I_2\otimes T_4$, and so on.

\item Constructing Symmetric Boolean Game:

Consider $G\in {\cal G}_{[4;2]}$, and denote
$$
x=(A,B,C,D,E,F,G,H)^T\in \R^8.
$$
Using Proposition \ref{p7.2.1} and equation (\ref{7.1.11}), we have that $G$ is an ordinary symmetric game, if and only if,
\begin{align}\label{7.2.4}
\begin{array}{ccl}
V^c_1&=&(H_4x)^T\\
~&=&[A,B,B,C, B,C,C,D, E,F,F,G, F,G,G,H];\\
V^c_2&=&V^c_1W_{[2,2]}\\
~&=&[A,B,B,C, E,F,F,G, B,C,C,D, F,G,G,H];\\
V^c_3&=&V^c_1W_{[4,2]}\\
~&=&[A,B,E,F, B,C,F,G, B,C,F,G, C,D,G,H];\\
V^c_4&=&V^c_1W_{[8,2]}\\
~&=&[A,E,B,F, B,F,C,G, B,F,C,G, C,G,D,H].\\
\end{array}
\end{align}
We conclude that $G\in {\cal G}_{[4;2]}$ is an ordinary symmetric game, if and only if,
\begin{align}\label{7.2.5}
\begin{array}{ccl}
V_G&=&AV_1+BV_2+CV_3+DV_4\\
~&~&+EV_5+FV_6+GV_6+HV_8,
\end{array}
\end{align}
where $\{V_i\;|\;i=1,2,\cdots,8\}$ form a basis of the ordinary symmetric subspace of ${\cal G}_{[4;2]}$, and they are:
$$
V_1=\begin{bmatrix}\d_{16}^1\\\d_{16}^1\\\d_{16}^1\\\d_{16}^1
\end{bmatrix}^T,
$$
$$
V_2=\begin{bmatrix}
\d_{16}^2+\d_{16}^3+\d_{16}^5\\
\d_{16}^2+\d_{16}^3+\d_{16}^9\\
\d_{16}^2+\d_{16}^5+\d_{16}^9\\
\d_{16}^3+\d_{16}^5+\d_{16}^9\\
\end{bmatrix}^T,
$$
$$
V_3=\begin{bmatrix}
\d_{16}^4+\d_{16}^6+\d_{16}^7\\
\d_{16}^4+\d_{16}^{10}+\d_{16}^{11}\\
\d_{16}^6+\d_{16}^{10}+\d_{16}^{13}\\
\d_{16}^7+\d_{16}^{11}+\d_{16}^{13}\\
\end{bmatrix}^T,
$$
$$
V_4=\begin{bmatrix}
\d_{16}^8\\
\d_{16}^{12}\\
\d_{16}^{14}\\
\d_{16}^{15}\\
\end{bmatrix}^T,
$$
$$
V_5=\begin{bmatrix}
\d_{16}^9\\
\d_{16}^{5}\\
\d_{16}^{3}\\
\d_{16}^{9}\\
\end{bmatrix}^T,
$$
$$
V_6=\begin{bmatrix}
\d_{16}^{10}+\d_{11}^6+\d_{16}^{13}\\
\d_{16}^6+\d_{16}^{7}+\d_{16}^{13}\\
\d_{16}^4+\d_{16}^{7}+\d_{16}^{11}\\
\d_{16}^4+\d_{16}^{6}+\d_{16}^{10}\\
\end{bmatrix}^T,
$$
$$
V_7=\begin{bmatrix}
\d_{16}^{12}+\d_{16}^{14}+\d_{16}^{15}\\
\d_{16}^8+\d_{16}^{14}+\d_{16}^{15}\\
\d_{16}^8+\d_{16}^{12}+\d_{16}^{15}\\
\d_{16}^8+\d_{16}^{12}+\d_{16}^{14}\\
\end{bmatrix}^T,
$$

$$
V_8=\begin{bmatrix}
\d_{16}^{16}\\
\d_{16}^{16}\\
\d_{16}^{16}\\
\d_{16}^{16}\\
\end{bmatrix}^T.
$$
\end{enumerate}
\end{exa}

Denote by ${\cal S}^o_{[n;\kappa]}\subset {\cal G}_{[n;\kappa]}$ the subspace of ordinary symmetric games. Then Theorem \ref{t7.1.4} and Proposition \ref{p7.2.1} imply the following result.

\begin{prp}\label{p7.2.3}
\begin{enumerate}
\item $G\in {\cal S}^o_{[n;2]}$, if and only if, the equation
\begin{align}\label{7.2.6}
\begin{bmatrix}I_{2^n}\\
W_{[2,2]}\otimes I_{2^{n-2}}\\
W_{[2,4]}\otimes I_{2^{n-3}}\\
\vdots\\
W_{[2,2^{n-1}]}
\end{bmatrix}H_nv=V_G^T
\end{align}
has solution $v\in \R^{2n}$.
\item
\begin{align}\label{7.2.7}
\dim\left({\cal S}^o_{[n;2]}\right)=2n.
\end{align}
\end{enumerate}
\end{prp}

\subsection{Symmetric Game vs Potential Game}%s-7.3

Applying (\ref{7.2.6}) to Theorem \ref{t5.1.2} yields the following result:

\begin{lem}\label{l7.3.1} A symmetric game $G\in {\cal S}^o_{[n;2]}$ is a potential game, if and only if, for any $v\in \R^{2n}$
 \begin{align}\label{7.3.1}
E(n)\xi(n)=B_nv
\end{align}
has solution $\xi$, where
 \begin{align}\label{7.3.2}
B(n)=\begin{bmatrix}
-I_{2^n}+W_{[2,2]}\otimes I_{2^{n-2}}\\
-I_{2^n}+W_{[2,4]}\otimes I_{2^{n-3}}\\
-I_{2^n}+W_{[2,8]}\otimes I_{2^{n-4}}\\
\vdots\\
-I_{2^n}+W_{[2,2^{n-1}]}
\end{bmatrix}.
\end{align}
\end{lem}

Note that since (\ref{7.3.1}) has solution for arbitrary $v$, equation (\ref{7.3.1}) is equivalent to
\begin{align}\label{7.3.3}
E(n)K_n=B(n)H_n
\end{align}
has solution $K_n\in {\cal M}_{n2^{n-1}\times 2n}$. In fact, $\Col_i(K_n)$ is the solution $\xi(n)$ for $v=\d_{2n}^i$, $i=1,2,\cdots,2n$.

\begin{lem}\label{l7.3.2} The solution $K_n$ of the equation (\ref{7.3.3}) exists.
\end{lem}
\noindent{\it Proof.} See the Appendix B.\hfill $\Box$

Splitting
$$
K_n=\begin{bmatrix}
K_n^1\\K_n^2\\
\vdots\\
K_n^n
\end{bmatrix},
$$
where $K_n^i\in {\cal M}_{2^{n-1}\times 2n}$, $i=1,\cdots,n$.
Then we have the following conclusion:

\begin{thm}\label{t7.3.3}
 Given a Boolean game $G\in {\cal G}_{[n;2]}$ with $V_G$ satisfying (\ref{7.2.6}), i.e., $G\in {\cal S}^o_{[n;2]}$ . Then it is a potential game. Moreover, the structure vector of its potential function is
\begin{align}\label{7.3.4}
V^P=v^T\left(H_n^T-(K_n^1)^TE_1^T\right),
\end{align}
where $v\in \R^{2n}$ is the solution of (\ref{7.2.6}).
\end{thm}

\begin{exa}\label{e7.3.4} Consider a $G\in {\cal S}^o_{[4;2]}$ as described in a general form (\ref{7.2.5}). It is easy to calculate that
$$
T_4=\begin{bmatrix}
     0&    -1&    -1&    -1&     1&     1&     1&     0\\
     0&     0&    -1&    -1&     0&     1&     1&     0\\
     0&     0&    -1&    -1&     0&     1&     1&     0\\
     0&     0&     0&    -1&     0&     0&     1&     0\\
     0&     0&    -1&    -1&     0&     1&     1&     0\\
     0&     0&     0&    -1&     0&     0&     1&     0\\
     0&     0&     0&    -1&     0&     0&     1&     0\\
     0&     0&     0&     0&     0&     0&     0&     0
\end{bmatrix}
$$
By definition
$$
E_1={\bf 1}_2\otimes I_{2^3}.
$$
And the $H_4$ is obtained in Example \ref{e7.2.2}. Using formula (\ref{7.3.4}),
we can calculate the structure vector of $G$ as
\begin{align}\label{7.3.7}
V^P=v^T\left(H_4^T-(K_4^1)^TE_1^T\right):=v^T\Psi,
\end{align}
where
$$
v=[A,B,C,D,E,F,G,H]^T\in \R^8,
$$
and
$$
\begin{array}{l}
\Psi=\\
\left[
\begin{array}{cccccccccccccccc}
1&0&0&0&0&0&0&0&0&0&0&0&0&0&0&0\\
1&1&1&0&1&0&0&0&1&0&0&0&0&0&0&0\\
1&1&1&1&1&1&1&0&1&1&1&0&1&0&0&0\\
1&1&1&1&1&1&1&1&1&1&1&1&1&1&1&0\\
-1&0&0&0&0&0&0&0&0&0&0&0&0&0&0&0\\
-1&-1&-1&0&-1&0&0&0&-1&0&0&0&0&0&0&0\\
-1&-1&-1&-1&-1&-1&-1&0&-1&-1&-1&0&-1&0&0&0\\
0&0&0&0&0&0&0&0&0&0&0&0&0&0&0&1
\end{array}\right].
\end{array}
$$
Note that (\ref{7.3.7}) provides a general formula for the potential function of $G\in {\cal S}^o_{[4;2]}$.
\end{exa}

\subsection{Weighted (Boolean) Games}%s-7.4

\begin{prp}\label{p7.4.1} Consider  a Boolean game $G\in {\cal G}_{[n;\kappa]}$. Assume $G$ is weighted symmetric satisfying (\ref{3.1.1}). Then $G$ is a weighted potential game with $w_i=\frac{1}{\mu_i}$, $i=1,\cdots,n$.
\end{prp}

\noindent{\it Proof}. Construct an auxiliary game $G^{\mu}$ by setting its payoff functions as
$$
c^{\mu}_i(x):=\mu_ic_i(x),\quad i=1,\cdots,n.
$$
Then (\ref{3.1.1}) implies that $G^{\mu}$ is an ordinary symmetric game. According to Theorem \ref{t7.3.3}, $G^{\mu}$ is a potential game. Hence there exists a potential function $P$ such that
\begin{align}\label{7.4.1}
\begin{array}{l}
\mu_ic_i(x_i,s_{-i})-\mu_ic_i(y_i,s_{-i})=P(x_i,s_{-i})-P(y_i,s_{-i}),\\
 i=1,\cdots,n.
\end{array}
\end{align}
(\ref{7.4.1}) implies that $G$ is a weighted potential game with $w_i=\frac{1}{\mu_i}$, $i=1,\cdots,n$.
\hfill $\Box$

Using $w_i=\frac{1}{\mu_i}$, (\ref{5.1.4}) and (\ref{5.1.6}) can be converted to
\begin{align}\label{7.4.2}
E_{\mu}(n):=\begin{bmatrix}
-\mu_1E_1&\mu_2E_2&0&\cdots&0\\
-\mu_1E_1&0&\mu_3E_3&\cdots&0\\
\vdots&~&~&~&~\\
-\mu_1E_1&0&0&\cdots&\mu_nE_n
\end{bmatrix}
\end{align}
and
\begin{align}\label{7.4.3}
b_{\mu}(n)=\begin{bmatrix}
b_{\mu}^2\\
b_{\mu}^3\\
\vdots\\
b_{\mu}^n
\end{bmatrix}=\begin{bmatrix}
\left(\mu_2 V^c_2-\mu_1V^c_1\right)^T\\
\left(\mu_3 V^c_3-\mu_1V^c_1\right)^T\\
\vdots\\
\left(\mu_n V^c_n-\mu_1V^c_1\right)^T
\end{bmatrix}.
\end{align}
It follows that (\ref{5.1.7}) becomes
\begin{align}\label{7.4.4}
E_{\mu}(n)\xi_{\mu}(n)=b_{\mu}(n).
\end{align}

Hence, we have

\begin{cor}\label{c7.4.2}
The potential function of weighted symmetric Boolean game $G\in {\cal S}^w_{[n;2]}$ is determined by its structure vector as
\begin{align}\label{7.4.5}
V^P=\mu_1\left(V^c_1-\left(\xi_{\mu}^1\right)^T E_1^T\right),
\end{align}
where $\xi_{\mu}^1\in \R^{2^{n-1}}$ is the first block of the solution of (\ref{7.4.4}).
\end{cor}

Finally, to verify whether $G\in {\cal G}_{[n;\kappa]}$ is weighted symmetric, the above argument leads Theorem \ref{t7.1.4} to the following modification.

\begin{prp}\label{t7.4.3} A game $G\in {\cal G}_{[n;\kappa]}$ is weighted symmetric with respect to $\{\mu_i, \; i=1,\cdots,n\}$, if and only if,
(\ref{7.1.10}) and the following (\ref{7.4.6}) hold with some $e_i=\frac{\mu_1}{\mu_i}>0$, $i=2,\cdots,n$.
\begin{align}\label{7.4.6}
V^c_i=e_iV^c_1W_{[\kappa^{i-1},\kappa]},\quad i=2,3,\cdots,n.
\end{align}
\end{prp}

Next, we consider a renaming symmetric game $G\in {\cal S}^r_{[n;\kappa]}$. If $\kappa=2$, then it is obvious that
$G$ is a potential game. Then we try to calculate its potential function. We ask a general question: When a renaming game is a potential game?

%{\color{green}
%Assume the renamed game $G^r$ has structure vectors $\tilde{V}^c_i$, $i=1,\cdots,n$. Then we have
%\begin{align}\label{7.4.7}
%\begin{array}{l}
%V^c_ix=\tilde{V}^c_i\ltimes_{j=1}^n\left(\Phi_{r_i^{-1}(j)}x\right)\\
%~~~=\tilde{V}^c_i\left(\Phi_{r_i^{-1}(1)}*\cdots*\Phi_{r_i^{-1}(n)}\right)x\\
%~~~:=\tilde{V}^c_i{\bf \Phi}_ix,
%\end{array}
%\end{align}
%where
%$$
%{\bf \Phi}_i=\Phi_{r_i^{-1}(1)}*\cdots*\Phi_{r_i^{-1}(n)}.
%$$
%That is,
%\begin{align}\label{7.4.8}
%\tilde{V}^c_i=V^c_i{\bf \Phi}_i^{-1},\quad i=1,\cdots,n.
%\end{align}
%
%Then we have the following
%
%\begin{prp}\label{p7.4.4} The renamed game $G^r$ is a potential game, if and only if,
%\begin{align}\label{7.4.9}
%E(n)\xi(n)=b_r(n)
%\end{align}
%has solution, where $E(n)$, $\xi(n)$ are as in (\ref{5.1.4}) and (\ref{5.1.5}) respectively with $w_i=1$ and
%\begin{align}\label{7.4.10}
%b_r(n)=
%\begin{bmatrix}
%b_2\\
%b_3\\
%\vdots\\
%b_n
%\end{bmatrix}=
%b_r(n)=
%\begin{bmatrix}
%(V^c_2{\bf \Phi}_2^{-1}-V^c_1{\bf \Phi}_1^{-1})^T\\
%(V^c_3{\bf \Phi}_3^{-1}-V^c_1{\bf \Phi}_1^{-1})^T\\
%\vdots\\
%(V^c_n{\bf \Phi}_n^{-1}-V^c_1{\bf \Phi}_1^{-1})^T
%\end{bmatrix}.
%\end{align}
%\end{prp}

Finally, from equation (\ref{7.4.1}) one sees that the potential function of a weighted symmetric game $G\in {\cal S}^w_{[n;\kappa]}$ is the same as its modified game $G^{\mu}$, which is a symmetric game. Hence if $G\in {\cal S}^w_{[n;2]}$, the formula (\ref{7.3.4}) can be used to calculate its potential function. We give an example to depict it.

\begin{exa}\label{e7.4.5}  Recall Example \ref{e3.1.3}. Since $\mu_1=3$ and $\mu_2=2$, its modified form $G^{\mu}$ has payoff bi-matrix as in Table \ref{Tab7.4}.

\begin{table}[!htbp] %Default value is [tbp]
\centering \caption{Payoff Bi-matrix of $G^{\mu}$\label{Tab7.4}}
%\vskip 2mm
\doublerulesep 0.5pt
\begin{tabular}{|c||c|c|}
\hline
\hline $P_1 \backslash P_2$&$1$&$2$\\
\hline $1$& $6,~6$&$12,~18$\\
\hline $2$&$18,~12$&$12,~12$ \\
\hline
\end{tabular}
\end{table}

Then
$$V_{G^{\mu}}=[6,12,18,12,6,18,12,12].$$
Using formula (\ref{7.3.7}),
$$
\begin{array}{ccl}
V_{G^{\mu}}&=&v^T[H_2^T,H_2^TW_{[2,2]}^T]\\
~&=&v^T\begin{bmatrix}
1&0&0&0&1&0&0&0\\
0&1&0&0&0&0&1&0\\
0&0&1&0&0&1&0&0\\
0&0&0&1&0&0&0&1
\end{bmatrix}.
\end{array}
$$
It follows that
$$
v=[6,12,18,12]^T.
$$
Note that $H_2=I_4$, $E_1={\bf 1}_2\otimes I_2$,
$$
K_2^1=
\begin{bmatrix}
0&0&1&0\\
0&1&0&0
\end{bmatrix}.
$$
Substituting them into (\ref{7.3.4}) yields
$$
V^P=v^T\Psi
$$
where
$$
\Psi=H_2^T-K_2^1E_1^T=
\begin{bmatrix}
1&0&0&0\\
0&0&0&-1\\
-1&0&0&0\\
0&0&0&1
\end{bmatrix}.
$$
Finally, we have
$$
V^P=[-12,0,0,0],
$$
and then the potential function is
$$
P(x_1,x_2)=[-12,0,0,0]x_1x_2.
$$
\end{exa}

\subsection{Renaming (Boolean) Games}%s-7.5

Consider a renaming symmetric game $G\in {\cal S}^r_{[n;\kappa]}$. Assume the renamed game $G^r$ has structure vectors $V^r_i$, $i=1,\cdots,n$. From the proof of Proposition \ref{p3.2.3}, we know
\begin{align}\label{7.5.2}
V^r_i=V^c_i \Gamma_r^T,\; i=1,\cdots,n.
\end{align}
where
$$
\Gamma_r:=P_{r_1}\ot P_{r_2}\ot\cdots\ot P_{r_n}.
$$

\begin{dfn}\label{d7.5.1} A game $G\in {\cal G}_{[n;\kappa]}$ is  renaming potential, if there exist a renaming $r=(r_1,\cdots,r_n)$, $r_i\in {\bf S}_{\kappa}$, $i=1,\cdots,n$ such that the renamed game $G^r$ is potential.
\end{dfn}

Note that $r_i\in S_{\kappa}$ changes only the order of the strategies within each $S_i$, so if the renaming game $G^r$ satisfies
\begin{align}\label{7.5.3}
\begin{array}{l}
c_i^r(x_i,s_{-i})-c_i^r(y_i,s_{-i})=P^r(x_i,s_{-i})-P^r(y_i,s_{-i}),\\
x_i,y_i\in S_i,\:s_{-i}\in S_{-i},\;i=1,\cdots,n,
\end{array}
\end{align}
then $G$ also satisfies (\ref{7.5.3}).

Applying Theorem \ref{t5.1.2} to $G^r$ yields the following:

\begin{prp}\label{p7.5.2}
\begin{enumerate}
\item A game $G\in {\cal G}_{[n;\kappa]}$ is renaming potential with renaming $r=(r_i\;|\; i=1,\cdots,n)$, if and only if,
\begin{align}\label{7.5.4}
E(n)\xi(n)=(I_{n-1}\ot \Gamma_r) b(n)
\end{align}
has solution, where $E(n)$, $b(n)$ and $\xi(n)$ are as in (\ref{5.1.5}) and (\ref{5.1.6}) respectively with $w_i=1$, $i=1,\cdots,n$.
\item If (\ref{7.5.4}) has solution, then the structure vector of a potential function is
\begin{align}\label{7.5.6}
V^{P^r}=V^c_1\Gamma_r^T -\xi_1^TE_1^T,
\end{align}
where $\xi_1$ is the first block of the solution of (\ref{7.5.4}).
\end{enumerate}
\end{prp}

Finally, we consider $G\in {\cal S}^e_{[n;2]}$. From the above argument we know that $G^r$ is ordinary symmetric and hence is potential. It follows that $G$ is also potential with the same potential function.
In addition to this, the formula (\ref{7.3.4}) can be used to calculate its potential function as follows: First, we construct the potential function for renamed system $G^r$
using (\ref{7.3.4}) as
$$
V^{P^r}=v^T\left(H_n^T-(K^1_n)^TE^T_1\right).
$$
Then we have the potential function with respect to $G^r$ as
$$
P^r(x)=V^{P^r}\ltimes_{i=1}^n x_i.
$$
Note that
$$
r_i^{-1}(x_i)=P_{r_i^{-1}}x_i=P_{r_i}^Tx_i,\; i=1,\cdots,n,
$$
and using Proposition \ref{p2.2.3}, we have
%{\color{blue}
%\begin{align}\label{7.5.6}
%\begin{array}{ccl}
%P(x)&=&P_r(x^r)=V^{P^r}\ltimes_{i=1}^nP_{r_i}^{-1}x_i\\
%~&=&V^{P^r}P_{r_1}^{-1}\left(I_2\otimes P_{r_2}^{-1}\right)
%\left(I_4\otimes P_{r_3}^{-1}\right)\\
%~&~&\cdots \left(I_{2^{n-1}}\otimes P_{r_n}^{-1}\right)\ltimes_{i=1}^nx_i\\
%~&:=&V^Px,
%\end{array}
%\end{align}
%where $x=\ltimes_{i=1}^nx_i$ and
%\begin{align}\label{7.5.7}
%V^P=V^{P^r}\ltimes_{i=1}^n \left(I_{i-1}\otimes P_{r_i}^{-1}\right).
%\end{align}
%}
\begin{align}\label{7.5.6}
\begin{array}{rl}
P^r(x)&= P(r_1^{-1}(x_1),\cdots, r_n^{-1}(x_n))\\
%&=V^P P_{r_1^{-1}} x_1\cdots P_{r_n^{-1}} x_n\\
&=V^P P_{r_1}^T\ot P_{r_2}^T\cdots \ot P_{r_n}^T x\\
&=V^P \Gamma_r^T x
\end{array}
\end{align}
and
\begin{align}\label{7.5.7}
V^P = V^{P^r} \Gamma_r.
\end{align}

We give an example to demonstrate it.

\begin{exa}\label{e7.5.3} Recall Example \ref{e3.2.2}. As we mentioned there, using $r_1=\id$ and $r_2=(1,2)$, we have a renamed system, which has its payoffs as in Table \ref{Tab7.5.1}

\begin{table}[!htbp] %Default value is [tbp]
\centering \caption{Payoff Bi-matrix of Renamed Battle of the Sexes\label{Tab7.5.1}}
%\vskip 2mm
\doublerulesep 0.5pt
\begin{tabular}{|c||c|c|}
\hline
\hline $P_1 \backslash P_2$&$F$&$C$\\
\hline $F$& $0,~0$&$2,~1$\\
\hline $C$&$1,~2$&$0,~0$ \\
\hline
\end{tabular}
\end{table}

Then we have
$$
V_{G^r}=[0,2,1,0,0,1,2,0].
$$

$$
\begin{array}{ccl}
V_{G^{r}}&=&v^T[H_2^T,H_2^TW_{[2,2]}^T]\\
~&=&v^T\begin{bmatrix}
1&0&0&0&1&0&0&0\\
0&1&0&0&0&0&1&0\\
0&0&1&0&0&1&0&0\\
0&0&0&1&0&0&0&1
\end{bmatrix}.
\end{array}
$$
It follows that
$$
v=[0,2,1,0]^T.
$$
Note that $H_2=I_4$, $E_1={\bf 1}_2\otimes I_2$,
$$
K_2^1=
\begin{bmatrix}
0&0&1&0\\
0&1&0&0
\end{bmatrix}.
$$
Substituting them into (\ref{7.3.4}) yields
$$
V^{P^r}=v^T\Psi
$$
where
$$
\Psi=H_2^T-(K_2^1)^T E_1^T=
\begin{bmatrix}
1&0&0&0\\
0&0&0&-1\\
-1&0&0&0\\
0&0&0&1
\end{bmatrix}.
$$

Hence
$$
V^{P^r}=v^T\Psi=[-1,0,0,-2].
$$
Using formula (\ref{7.5.7}) yields
$$
V^P=V^{P^r}(I_2\otimes M_n)=[0,-1,-2,0].
$$
Hence the potential function of the Battle of the Sexes is
$$
P(x)=[0,-1,-2,0]x_1x_2.
$$
\end{exa}

\section{Negation-symmetric Boolean Games}%s-8

In previous sections it was seen that a (ordinary or renaming or weighted) symmetric Boolean game is a potential game (could be weighted one). Carefully verifying Example \ref{e4.12} can show that in general a name-irrelevant symmetric game $G\in {\cal G}_{[n;2]}$ is not necessarily to be a (weighted) potential game. This section shows that a special class of $G\in {\cal G}_{[n;2]}$, which is not a class of (ordinary or renaming or weighted) symmetric Boolean game, is also potential.

\begin{dfn}\label{d8.1} A game
$G\in {\cal G}_{[n;2]}$ is called a negation-symmetric Boolean game, if
\begin{align}\label{8.1}
\begin{array}{ccl}
c_i(\neg x_1,x_2,\cdots,\neg x_i,\cdots,x_n)&=&c_1(x_1,\cdots,x_n),\\
~&~& i=2,\cdots,n.
\end{array}
\end{align}
\end{dfn}

The following proposition can be used to verify if $G$ is negation-symmetric.

\begin{prp}\label{p8.2} A game $G\in {\cal G}_{[n;2]}$ is negation-symmetric, if and only if,
\begin{align}\label{8.2}
V^c_i=V^c_1\left(I_{2^{i-1}}\otimes M_n\right)M_n,\quad i=2,3,\cdots,n.
\end{align}
\end{prp}

\noindent{\it Proof}. Express (\ref{8.1}) into algebraic form, and then use Proposition \ref{p2.2.3}, (\ref{8.2}) can easily be obtained.
\hfill $\Box$

\begin{exa}\label{e8.3}
Given a game $G\in {\cal G}_{[2;2]}$. $G$ is a negation-symmetric game, if and only if, its payoff bi-matrix is as in Table \ref{Tab8.1}.

\begin{table}[!htbp] %Default value is [tbp]
\centering \caption{Payoff Bi-matrix \label{Tab8.1}}
%\vskip 2mm
\doublerulesep 0.5pt
\begin{tabular}{|c||c|c|}
\hline
\hline $P_1 \backslash P_2$&$1$&$2$\\
\hline $1$& $a,~b$&$c,~d$\\
\hline $2$&$d,~c$&$b,~a$ \\
\hline
\end{tabular}
\end{table}
\end{exa}

\begin{thm}\label{t8.4} A negation-symmetric Boolean game is a potential game.
\end{thm}

\noindent{\it Proof.} Using (\ref{8.2}), we have
\begin{align}\label{8.3}
\begin{array}{ccl}
\left(V^c_i\right)^T -\left(V^c_1\right)^T&=&\left[M_n\left(I_{2^{i-1}}\otimes M_n\right)-I_{2^i}  \right]\left(V^c_1\right)^T,\\
~&~& i=2,\cdots,n.
\end{array}
\end{align}
From (\ref{8.3}), an algebraic computation shows that
\begin{align}\label{8.4}
\begin{array}{ccl}
\begin{bmatrix}
\left(V^c_2-V^c_1\right)^T\\
\left(V^c_3-V^c_1\right)^T\\
\vdots\\
\left(V^c_n-V^c_1\right)^T\\
\end{bmatrix}&=&
\begin{bmatrix}
M_n\otimes M_n\otimes I_{2^{n-2}}-I_{2^n}\\
M_n\otimes I_2\otimes M_n\otimes I_{2^{n-3}}-I_{2^n}\\
\vdots\\
M_n\otimes I_{2^{n-2}}\otimes M_n-I_{2^n}\\
\end{bmatrix}\left(V^c_1\right)^T\\
~&:=&\Gamma\left(V^c_1\right)^T.
\end{array}
\end{align}
Substituting it into the potential equation (\ref{5.1.7}) with $w_i=1$ leads to
\begin{align}\label{8.5}
E(n)B=\Gamma,
\end{align}
where
\begin{align}\label{8.6}
B=\begin{bmatrix}
{\bf 1}_2^T\otimes I_{2^{n-1}}\\
M_n\otimes {\bf 1}_2^T\otimes I_{2^{n-2}}\\
M_n\otimes I_2\otimes {\bf 1}_2^T\otimes I_{2^{n-3}}\\
\vdots\\
M_n\otimes I_{2^{n-2}}\otimes {\bf 1}_2^T
\end{bmatrix}.
\end{align}

(We refer the readers to Appendix C for details.)

It follows that
\begin{align}\label{8.7}
B\left(V^c_1\right)^T
\end{align}
is a solution of the potential equation (\ref{5.1.7}) (with $w_i=1$ $\forall i$) .
The conclusion follows.
\hfill $\Box$

It is easy to figure out that
$$
\xi_1=\left[{\bf 1}_2^T\otimes I_{2^{n-1}}\right]\left(V^c_1\right)^T.
$$
It follows from (\ref{5.1.8}) that the potential function has the structure vector as
\begin{align}\label{8.8}
\begin{array}{ccl}
V^P&=&V^c_1\left[I_{2^n}-{\bf 1}_2\otimes I_{2^{n-1}}E_1^T\right]\\
~&=&-V^c_1\left[M_n\otimes I_{2^{n-1}}\right].
\end{array}
\end{align}

\section{Concluding Remarks}

This paper considers the symmetry of finite games $G\in {\cal G}_{[n;\kappa]}$. Roughly speaking, it was shown that $G$ is symmetric with respect to a $\theta\in {\bf \Theta}_{[n;\kappa]}\subset {\bf S}_{n\kappa}$ means the structure vector $V_G$ is invariant with respect to a linear representation of $\theta$.

More precisely, the linear representation of $\theta$ can be characterized as
$$
\left(P_{\pi_{\theta}}; D^{\theta}_{1},D^{\theta}_{2},\cdots, D^{\theta}_{n}\right).
$$
Particularly, this representation can be classified as follows:
\begin{itemize}
\item When $ D^{\theta}_{i}=I_{\kappa}$, $\forall i$, $G$ is ordinary symmetric.
\item When $ D^{\theta}_{i}=w_iI_{\kappa}$, $G$ is weighted symmetric.
\item When  $ D^{\theta}_{i}=D_i$, (i.e., it is independent of $\theta$) $G$ is renaming symmetric.
\item For general  $D^{\theta}_{i}$, $G$ is strategy-permutation symmetric with respect to $\theta$.
And as $\Pi(G)={\bf S}_n$, $G$ is name-irrelevant symmetric.
\end{itemize}
The linear representation of ${\cal S}_{n\kappa}$ (or its subgroups) in the structure vector space $V_{G}$ of $G$ is a convenient tool to investigate the properties of symmetric games.

Then the relationship between symmetric games and potential games is investigated. It was shown that $G\in {\cal S}^o_{[n;2]}$ and $G\in {\cal S}^r_{[n;2]}$ are also potential games. $G\in {\cal S}^w_{[n;2]}$ is also a weighted potential game. The formulas for calculating their corresponding potential functions are also presented. In addition to this three kind of Boolean games, a class of $G\in {\cal G}_{[n;2]}$, called the negation-symmetric Boolean games, is also proved to be potential.

Several problems remain for further investigation:
\begin{enumerate}
\item General dimension and basis of ${\cal S}^o_{[n;\kappa]}$ are still unknown.
\item When a networked game is symmetric?
\item Orthogonal decomposition of ${\cal G}_{[n;\kappa]}$ into ${\cal S}^o_{[n;\kappa]}$ and $({\cal S}^o_{[n;\kappa]})^{\perp}$ is still unknown.
\item More properties of symmetric games may be deduced from their linear representations.
\end{enumerate}

\vskip 5mm

\noindent{\bf Appendix}

\subsection*{ A. The proof of Proposition \ref{p7.2.1}:}
We give a lemma first.
\begin{lem}\label{cl}
Let $h_i=\Col_i(H_n)$, $i=1,\cdots,2n$. Then we have
\begin{align}\label{cl.1}
h^T_i x=
\begin{cases}
1,& x_1=\d_2^1,\; \sharp(x_{-1},1) = n-i, \:i\leq n,\\
1,& x_1=\d_2^2,\;\sharp(x_{-1},1) = 2n-i,\: i>n,\\
0,& \text{Otherwise.}
\end{cases}
\end{align}
\end{lem}

\noindent{\it Proof}. Form the definition of $H_n$ in (\ref{7.2.3}), $H_n=I_2\ot T_n$ where $T_n\in {\cal B}_{2^{n-1}\times n}$. Denote the $i$-th column of $T_n$ as $t_i$, then we have
\begin{align}\label{cl.2}
h_i^T=
\begin{cases}
\left[t_{i}^T, {\bf 0}_{2^{n-1}}^T\right],& i=1,\cdots,n,\\
\left[{\bf 0}_{2^{n-1}}^T, t_{i-n}^T\right], & i=n+1,\cdots,2n.
\end{cases}
\end{align}
Substituting (\ref{cl.2}) into (\ref{cl.1}), we claim that (\ref{cl.1}) holds, if and only if,
\begin{align}\label{cl.3}
t_i^T x_2\cdots x_n=
\begin{cases}
1,& \sharp(x_{-1},1) = n-i,\\
0,& \text{Otherwise.}
\end{cases}\; i=1,\cdots,n.
\end{align}
 The claim can be proved using mathematical induction: It is easy to verify (\ref{cl.3}) for $n=2$. Then we set $t_i=\Col_i(T_k)$ and assume (\ref{cl.3}) holds for $n=k$. That is,
\begin{align}\label{cl.5}
t_i^T x_2\cdots x_k=
\begin{cases}
1,&\sharp(x_{-1},1) = k-i,\\
0,&\text{Otherwise.}
\end{cases}\; i=1,\cdots,k.
\end{align}

From the definition of $T_{k+1}$ as in (\ref{7.2.2}), we have
\begin{align}\label{cl.6}
T_{k+1}=\begin{bmatrix}t_1&t_2& \cdots &t_k & t_{k+1}\\
t_0 & t_1&\cdots & t_{k-1} &t_k
\end{bmatrix},
\end{align}
where $t_0=t_{k+1}={\bf 0}_{2^{k-1}}$.

Then the $s$-th column of $T_{k+1}$ is $[t_s^T, t_{s-1}^T]^T$. Using assumption (\ref{cl.5}), we get
\begin{align}\label{cl.7}
\begin{array}{ccl}
&~&[t_s^T, t_{s-1}^T]x_2x_3\cdots x_{k+1}\\
&=&[t_s^Tx_3\cdots x_{k+1}, t_{s-1}^Tx_3\cdots x_{k+1}]x_2\\
&=&\begin{cases}
1 & x_2=\d_2^1 ,\;\sharp(x_{-\{1,2\}},1)= k-s, \\
1 & x_2=\d_2^2,\; \sharp(x_{-\{1,2\}},1)= k-s+1, \\
0 & \text{Otherwise.}
\end{cases}\\
&=&\begin{cases}
1 & \sharp(x_{-1},1)= k+1-s, \\
0 & \text{Otherwise.}
\end{cases}
\end{array}
\end{align}
This implies that (\ref{cl.3}) is true for $n=k+1$. Therefore, (\ref{cl.3}) holds for every $n$ and (\ref{cl.1}) follows. 
\hfill $\Box$

Now it is ready to prove Proposition \ref{p7.2.1}.

\noindent{\it Proof}.
Proposition \ref{p7.2.1} is equivalent to that the column set of $H_n$ contains a basis of $V_1^c$ for any symmetric boolean game $G\in {\cal S}^o_{[n;2]}$.

From the argument before Proposition \ref{p7.2.1}, we know that the dimension of $V_1^c$ for $G\in {\cal S}^o_{[n;2]}$ is $2n$. It is easy to check that $|\Col(H_n)|=2n$. Denote
\[
h_i=\Col_i(H_n),\; i=1,\cdots, 2n.
\]
Then, it is sufficient show that $\{h_i, i=1,\cdots,2n\}$ is a basis of $V_1^c$ for ${\cal S}^o_{[n;2]}$.

First, we prove that $h_i$, $i=1,\cdots,2n$, are linearly independent.  To this end, let
\begin{align}\label{c.1.1}
\sum_{i=1}^{2n} a_i h_i = {\bf 0}_{2^n}.
\end{align}
Taking transpose on both sides and multiplying both sides by $x^i\in \D_{2^n}$ with $\sharp(x^i,1)=n-i$ for $i=1,\cdots n$ and $x^i$ with $\sharp(x^i,1)=2n-i$ for $i=n+1,\cdots, n$, then we have $a_i=0$, $i=1,\cdots,2n$. As a result, $h_i$, $i=1,\cdots,2n$ are linearly independent.

Second, we prove that each $h_i$ is a suitable candidate of $V_1^c$ for $G\in {\cal S}^o_{[n;2]}$. According to Proposition \ref{p7.1.2}, this means that
\begin{align}\label{c.1.4}
\begin{array}{l}
h_i^T x_1\cdots x_n = h_i^T y_1\cdots y_n,
\end{array}
\end{align}
where $x_1=y_1$, and $\sharp(x_{-1})=\sharp(y_{-1})$. Using Lemma \ref{cl}, it is straightforward to verify this.

Therefore, $\{h_i, i=1,\cdots,2n\}$ is a basis of $V_1^c$ for $G\in {\cal S}^o_{[n;2]}$, and Proposition \ref{p7.2.1} follows.
\hfill $\Box$

\subsection*{ B. The proof of Lemma \ref{l7.3.2}:}

We prove it by construction. Set
\begin{align}\label{a.1.1}
K_n:={\bf 1}_n\otimes Q_n,
\end{align}
where
\begin{align}\label{a.1.2}
Q_n:=[{\bf 0}_{2^{n-1}},-R_n,R_n, {\bf 0}_{2^{n-1}}],
\end{align}
and $R_n\in {\cal M}_{2^{n-1}\times (n-1)}$ is defined recursively as
\begin{align}\label{a.1.3}
\begin{cases}
R_2=\d_2^1,\\
R_{t+1}=\begin{bmatrix}
R_t&{\bf 1}_{2^{t-1}}\\{\bf 0}_{2^{t-1}}&R_t
\end{bmatrix},\quad t\geq 2.
\end{cases}
\end{align}
We prove that the $K_n$ defined by (\ref{a.1.1})-(\ref{a.1.3}) satisfies equation (\ref{7.3.3}). That is, we need to show that
\begin{align}\label{a.1.4}
\begin{array}{l}
(-E_1+E_{i})Q_n=\left(-I_{2^{n}}+W_{[2,2^{i-1}]}\otimes I_{2^{n-i}}\right)H_n.\\
~~~~~~~\qquad i=2,\cdots,n.
\end{array}
\end{align}

To this end, the following lemma is needed.
\begin{lem}\label{al}
Let $r_i=\Col_i(R_n)$, $i=1,\cdots,n-1$, then we have
\begin{align}\label{al.1}
r^T_i x_1\cdots x_{n-1}=
\begin{cases}
1,& \sharp(x,1)\geq n-i,\\
0,& \text{Otherwise.}
\end{cases}
\end{align}
\end{lem}
\noindent{\it Proof}.
We prove it by mathematical induction. It is easy to verify (\ref{al.1}) for $n=2$. Then we assume that (\ref{al.1}) holds for $n=k$. Denoting $r_i$ as the $i$-th column of $R_k$, we have
\begin{align}\label{al.2}
r^T_i x_1\cdots x_{k-1}=
\begin{cases}
1,& \sharp(x,1)\geq k-i,\\
0,& \text{Otherwise.}
\end{cases}
\end{align}
Next, we express $R_{k+1}$ row by row as
\begin{align}\label{al.3}
R_{k+1}=\begin{bmatrix}
r_1,& r_2 & \cdots & r_{k-1}& r_k\\
r_0& r_1,& \cdots &r_{k-2}& r_{k-1}\\
\end{bmatrix}
\end{align}
where $r_0={\bf 0}_{2^{k-1}}$ and $r_k={\bf 1}_{2^{k-1}}$.

Then the $s$-th column of $R_{k+1}$ is $[r_s^T, r_{s-1}^T]^T$.  Using assumption (\ref{al.2}), we have
\begin{align}
\label{al.4}
\begin{array}{ccl}
&~&[r_s^T, r_{s-1}^T]x_1x_2\cdots x_k\\
&=&[r_s^Tx_2\cdots x_k, r_{s-1}^Tx_2\cdots x_k]x_1\\
&=&\begin{cases}
1 & x_1=\d_2^1,\;\sharp(x_{-1},1)\geq k-s,\\
1 & x_1=\d_2^2,\; \sharp(x_{-1},1)\geq k-s+1,\\
0 & \text{Otherwise}
\end{cases}\\
&=&\begin{cases}
1 & \sharp(x,1)\geq k+1-s,\\
0 & \text{Otherwise}
\end{cases}
\end{array}
\end{align}
Therefore, (\ref{al.1}) holds for $n=k+1$ and the Lemma is proved.
\hfill $\Box$
Now it is ready to give the proof of Lemma \ref{l7.3.2}.
\noindent{\it Proof}.
Denote the $s$-th column of $Q_n$ and $H_n$ by $q_s$ and $h_s$ respectively. Substituting them into each row-block equations of (\ref{a.1.4}) and taking  transpose on both sides, we have the following equalities for $i=2,\cdots,n$:
\begin{align}\label{a.1.7}
\begin{array}{l}
q_s^T (-E_1^T + E_i^T) = h_s^T (-I_{2^n} + W_{[2^{i-1},2]}),\\
~~~~~~~~\qquad s=1,2,\cdots,2n.
\end{array}
\end{align}

Based on the fact that (\ref{a.1.7}) holds, if and only if, for any $x=\od_{i=1}^n x_i \in \D_{2^n}$,
\begin{align}\label{a.1.8}
\begin{array}{l}
q_s^T (-E_1^T + E_i^T) x = h_s^T (-I_{2^n} + W_{[2^{i-1},2]})x,\\
x=\od_{i=1}^n x_i \in \D_{2^n},\; s=1,2,\cdots,2n.
\end{array}
\end{align}
And equivalently,
\begin{align}\label{a.1.9}
\begin{array}{l}
q_s^T (-x_{-1} + x_{-i}) = h_s^T (-x_1 x_{-1} + x_i x_{-i}),\\
x=\od_{i=1}^n x_i \in \D_{2^n},\;  s=1,2,\cdots,2n.
\end{array}
\end{align}
Note that
\begin{align}\label{a.1.10}
q_s=\begin{cases}
-r_{s-1}, & s\leq n,\\
r_{s-n}, & s > n,
\end{cases}
\end{align}
where
\begin{align}\label{a.1.11}
r_s=\begin{cases}
\Col_s(R_n), & s=1,\cdots,n-1,\\
{\bf 0}_{2^{n-1}}, & s=0,\;n.
\end{cases}
\end{align}

If $x_1= x_i$, then $\sharp(x_{-i})=\sharp (x_{-1})$.  According to Lemma \ref{cl} and Lemma \ref{al}, it is easy to check that both sides of (\ref{a.1.8}) are equal to zero.

If $x_1\neq x_i$, without loss of generosity we assume that $x_1=\d_2^1$, and $x_i=\d_2^2$, then
\begin{align}\label{a.2.1}
\sharp(x_{-1},1)=\sharp(x_{-i},1)-1.
\end{align}
From (\ref{cl.1}), we know that the right-hand side (RHS) of (\ref{a.1.8}) equals to zero except
\begin{align}\label{a.2.2}
\begin{cases}
RHS=-1, & \sharp(x_{-1})= n-s, \: s\leq n,\\
RHS= 1,& \sharp(x_{-i})=2n-s, \: s>n,
\end{cases}
\end{align}
From (\ref{al.1}), we know that the left-hand side (LHS) of (\ref{a.1.8}) equals to zero except $LHS=1$ if
\begin{align}\label{a.2.3}
\begin{cases}
\sharp(x_{-1})\geq n-s+1 & \sharp(x_{-i},1)<n-s+1, s\leq n,\\
\sharp(x_{-i})\geq 2n-s, & \sharp(x_{-1},1)<2n-s, s>n,
\end{cases}
\end{align}
and $LHS=-1$ if
\begin{align}\label{a.2.4}
\begin{cases}
\sharp(x_{-1})< n-s+1,&\sharp(x_{-i},1)\geq n-s+1,\: s\leq n,\\
\sharp(x_{-i})< 2n-s,& \sharp(x_{-1},1)\geq 2n-s,\: s>n.
\end{cases}
\end{align}

Taking the constraint (\ref{a.2.1}) into consideration, then
\begin{align}\label{a.2.5}
\begin{cases}
LHS=-1, & \sharp(x_{-1},1)= n-s, \: s\leq n,\\
LHS= 1,& \sharp(x_{-i},1)=2n-s, \: s>n,\\
LHS=0, & \text{Otherwise.}
\end{cases}
\end{align}
Hence $LHS=RHS$.

Then (\ref{a.1.7}) follows and this completes the proof of Lemma \ref{l7.3.2}.
\hfill $\Box$

\subsection*{C. The proof (\ref{8.5}):}
For any $i=2,\cdots,n$, we have
\begin{align}\label{b.1.1}
\begin{array}{rl}
&-E_1 B_1 + E_i B_i\\
=&-E_1 {\bf 1}_2^T\ot I_{2^{n-1}} + E_i M_n\ot I_{2^{i-2}}\ot {\bf 1}^T_2\ot I_{2^{n-i}} \\
=&-\begin{bmatrix}1,1\\
1,1
\end{bmatrix}\ot I_{2^{n-1}} + M_n\ot I_{2^{i-2}}\ot \begin{bmatrix}1,1\\
1,1
\end{bmatrix}\ot I_{2^{n-i}} \\
=&-(M_n+I_2)\ot I_{2^{n-1}}+M_n\ot I_{2^{i-2}}\ot (M_n+I_2)\ot I_{2^{n-i}} \\
=&M_n\ot I_{2^{i-2}}\ot M_n\ot I_{2^{n-i}}-I_{2^n}\\
=& \Gamma_i,
\end{array}
\end{align}
where $B_i\in {\cal B}_{{2^{n-1}}\times 2^n}$ and $\Gamma_i\in {\cal B}_{2^n\times 2^n}$ are the $i$-th row block of $B$ and $\Gamma$ respectively.
Therefore, (\ref{8.5}) follows.
\hfill $\Box$

\end{document}